\documentclass[10pt,article,apl,aps,prx,superscriptaddress,amsmath,amssymb,floatfix,twocolumn,tightenlines]{revtex4-2}

\usepackage{amssymb}
\usepackage{natbib}

\usepackage{multirow}
\usepackage{bm}
\usepackage{amsmath}
\usepackage{graphicx}
\usepackage{epstopdf}
\usepackage{subfigure}
\usepackage{natbib}
\usepackage{epsfig}
\usepackage{amsfonts}
\usepackage{mathrsfs}
\usepackage{braket}
\usepackage{tabularx}
\usepackage{xcolor}
\usepackage[toc,page,title,titletoc,header]{appendix}
\usepackage{hyperref}
\hypersetup{colorlinks,allcolors=blue}

\usepackage{dsfont,amsthm,amsbsy}

\usepackage{fancyhdr}
\usepackage{lineno}
\begin{document}

\title{Link between thermodynamic correlation signatures and superconductivity \\ in twisted trilayer graphene}

\author{Jesse C. Hoke}
\email{jchoke@alumni.stanford.edu}
\altaffiliation{Current Address: IBM Quantum, San Jose, CA, USA}
\affiliation{Department of Physics, Stanford University, Stanford, CA 94305, USA}
\affiliation{Geballe Laboratory for Advanced Materials, Stanford, CA 94305, USA}
\affiliation{Stanford Institute for Materials and Energy Sciences, SLAC National Accelerator Laboratory, Menlo Park, CA 94025, USA}

\author{Yifan Li}
\affiliation{Department of Physics, Stanford University, Stanford, CA 94305, USA}
\affiliation{Geballe Laboratory for Advanced Materials, Stanford, CA 94305, USA}
\affiliation{Stanford Institute for Materials and Energy Sciences, SLAC National Accelerator Laboratory, Menlo Park, CA 94025, USA}

\author{Yuwen Hu}
\affiliation{Department of Physics, Stanford University, Stanford, CA 94305, USA}
\affiliation{Geballe Laboratory for Advanced Materials, Stanford, CA 94305, USA}
\affiliation{Stanford Institute for Materials and Energy Sciences, SLAC National Accelerator Laboratory, Menlo Park, CA 94025, USA}

\author{Julian May-Mann}
\affiliation{Department of Physics, Stanford University, Stanford, CA 94305, USA}

\author{Kenji Watanabe}
\affiliation{Research Center for Electronic and Optical Materials, National Institute for Materials Science, 1-1 Namiki, Tsukuba 305-0044, Japan}

\author{Takashi Taniguchi}
\affiliation{Research Center for Materials Nanoarchitectonics, National Institute for Materials Science,  1-1 Namiki, Tsukuba 305-0044, Japan}

\author{Trithep Devakul}
\affiliation{Department of Physics, Stanford University, Stanford, CA 94305, USA}

\author{Aaron Sharpe}
\email{aaron.sharpe@stanford.edu }
\affiliation{Department of Physics, Stanford University, Stanford, CA 94305, USA}
\affiliation{Stanford Institute for Materials and Energy Sciences, SLAC National Accelerator Laboratory, Menlo Park, CA 94025, USA}

\author{Benjamin E. Feldman}
\email{bef@stanford.edu}
\affiliation{Department of Physics, Stanford University, Stanford, CA 94305, USA}
\affiliation{Geballe Laboratory for Advanced Materials, Stanford, CA 94305, USA}
\affiliation{Stanford Institute for Materials and Energy Sciences, SLAC National Accelerator Laboratory, Menlo Park, CA 94025, USA}


\begin{abstract}
Twisted graphene multilayers exhibit strong electronic correlations leading to a range of experimental signatures. Yet how these signatures relate to each other and to the microscopic ground states---and how twist angle and band structure reshape them---remains poorly understood. Here we study this interplay by correlating local thermodynamic and transport measurements in twisted trilayer graphene (TTG) with unequal angles and flat electronic bands. We use a scanning single-electron transistor to map the impact of electron-electron interactions in the sample by selecting a region with smooth local twist angle evolution. We observe gapped correlated insulator states and asymmetric oscillations in the inverse electronic compressibility, both exhibiting pronounced electron-hole ($e$-$h$) asymmetry with distinct “magic” angles for conduction and valence bands. Subsequent transport measurements in the same region reveal robust superconductivity with a similar $e$-$h$ asymmetry. Our measurements indicate that superconductivity is not directly tied to the correlated insulator states. Instead, its critical temperature correlates closely with the strength of the compressibility oscillations, suggesting a common origin or link between the two. By combining a local probe with transport measurements, we uncover connections between superconductivity and thermodynamic correlation signatures that are not apparent from either technique in isolation. These findings highlight the power of our dual approach.

\end{abstract}

\maketitle

\begin{figure*}[t!]
    \renewcommand{\thefigure}{\arabic{figure}}
    \centering
    \includegraphics[width=1.99\columnwidth]{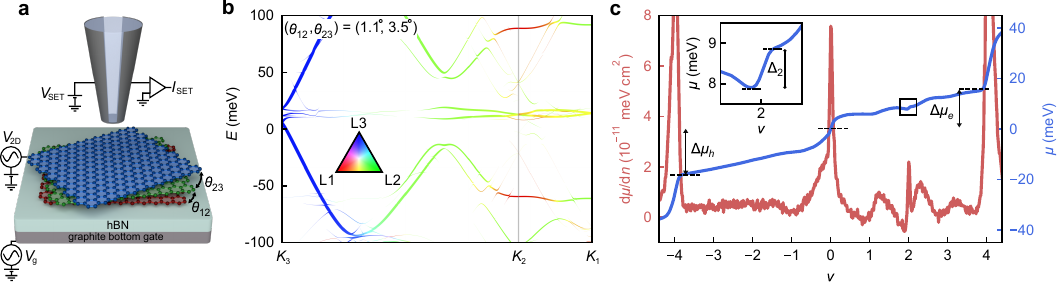}
    \caption{\textbf{Twisted trilayer graphene (TTG) with $\mathbf{\theta_{12}/\theta_{23} \approx 1/3}$}. \textbf{a}, Schematic of the measurement setup and device (Methods). Three graphene layers are sequentially twisted in the same direction (helical stacking), with a twist angle ratio near $\theta_{12}/\theta_{23}=1/3$. \textbf{b}, Calculated spectral function of TTG along a path between the $K$ points of each layer for $(\theta_{12}, \theta_{23}) = (1.1^\circ, 3.5^\circ)$. \textbf{c}, Inverse electronic compressibility d$\mu$/d$n$ (red) and chemical potential $\mu$ (blue) as a function of the moiré filling factor $\nu$ associated with layers 1 and 2 at a location where $\theta_{12} = 1.10^\circ$ and measured at temperature $T = 330$ mK. We estimate $\theta_{23} \approx 3.45^\circ$ at this location by measuring the renormalized Fermi velocity $v^*_F$ of the Dirac-like cone (Extended Data Fig. 2; Methods). $\Delta \mu_{e(h)}$ is defined as the total change in the chemical potential to fill the conduction (valence) bands. Inset: The step in $\mu$ at $\nu = 2$ defines the gap size $\Delta_2$ of the correlated insulator.} 
    \label{fig:fig1}
\end{figure*} 

Understanding how electron-electron interactions generate competing and coexisting orders is a longstanding challenge in condensed matter physics. Due to their high degree of tunability, moiré flat bands~\cite{Andrei_2020,Balents_2020,Mak_2022} are an excellent testbed for realizing and probing correlated states. The first discovered example, magic-angle twisted bilayer graphene (MATBG)~\cite{bistritzer2011moire}, hosts a wide range of interaction-driven effects, including correlated insulators~\cite{cao2018correlated,yankowitz2019tuning,lu2019superconductors,yu_spin_2022}, superconductivity~\cite{cao2018unconventional,yankowitz2019tuning,lu2019superconductors,arora2020superconductivity,cao2021nematicity}, and oscillatory compressibility as a function of filling~\cite{zondiner2020cascade,yu2022correlated}.
Transport experiments across different devices have used Coulomb screening to explore the relationship between superconductivity and the correlated insulators, finding that superconductivity could persist even in the absence of proximate correlated insulating states, or vice versa~\cite{saito2020independent,stepanov2020untying,liu2021tuning,gao2024double,barrier2024coulomb,wang2025independently}. However, a detailed understanding of the origin of superconductivity and its relation to other correlated ground states 
remains elusive.
Establishing the link between different correlation signatures, including those seen via other means such as thermodynamic measurements, is an important goal but has been hindered by sample-to-sample variability in twist angle, dielectric environment, strain, and/or disorder.

The same signatures of strong correlations are commonly found in other twisted graphene multilayers~\cite{park2021tunable,hao2021electric,liu2022isospin,cao2021pauli,zhang2022promotion,park2022robust,uri2023superconductivity,pierce2025tunable,chen2020tunable, zhang2021correlated}. 
These structures both expand the accessible parameter space and enable further modification of the electronic structure by the interference between multiple moiré patterns, providing new opportunities to probe strongly correlated phenomena with enhanced tunability.
The simplest example is twisted trilayer graphene (TTG): three graphene sheets with two independent interlayer twist angles $\theta_{12}$ and $\theta_{23}$ (here we assume $|\theta_{12}| \le |\theta_{23}|$ without loss of generality). Calculations predict that TTG hosts a ``magic continuum" of angle combinations exhibiting flat bands~\cite{zhu2020twisted, yang2024multi,foo2024extended,popov2023magic}. Along this continuum, the resulting correlated phenomena depend sensitively on the magnitude and relative sign (helical vs. alternating) of the twist angles.
Aside from work on mirror-symmetric magic-angle twisted trilayer graphene (MATTG; $\theta_{12} =- \theta_{23} \approx 1.55^\circ$)~\cite{park2021tunable,hao2021electric,liu2022isospin,cao2021pauli,banerjee2025superfluid,tanaka2025superfluid,pierce2025tunable,xie2025strong,turkel2022orderly,kim2022evidence,kim2023imaging,shen2023dirac}, only a handful of studies have explored other twist angle combinations along the magic continuum~\cite{uri2023superconductivity,hoke2024uncovering,xia2025topological,hoke2024imaging}.
The broader TTG phase diagram offers opportunities to probe how varying twist angles and layer numbers influence signatures of correlations---revealing whether the same generic features persist and any correspondence between them.

Here, we report local thermodynamic and charge transport measurements within a TTG sample that lies near the magic continuum with mismatched twist angles $\theta_{12}/\theta_{23} \approx 1/3$ (Fig.~\ref{fig:fig1}\textbf{a}). 
Electronic compressibility measurements with a single-electron transistor (SET) microscope show correlated insulators and oscillations in the compressibility that are maximized at different interlayer twists for the conduction and valence bands.
Transport measurements in the same region reveal robust superconducting phases with an $e$-$h$ asymmetry in the critical temperature that closely mirrors the strength of the compressibility oscillations, but not the correlated insulators. 
Our work provides a direct comparison of thermodynamic and transport signatures, including their joint evolution with twist angle, while avoiding confounding factors inherent in cross-device studies. 
We shed new light on the relationship between superconductivity and other correlation effects, simultaneously clarifying which structural and electronic parameters dominate electronic behavior in twisted graphene multilayers. 

\begin{figure*}[t!]
    \renewcommand{\thefigure}{\arabic{figure}}
    \centering
    \includegraphics[width=1.5\columnwidth]{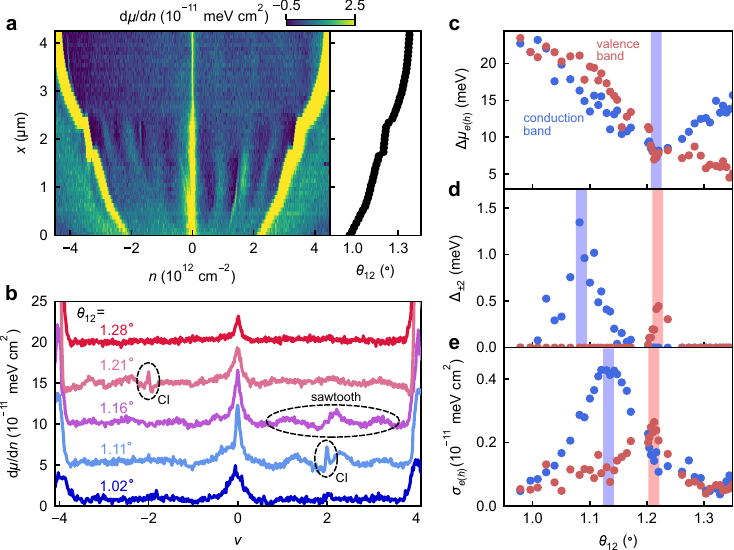}
    \caption{\textbf{Mapping twist-dependent interaction strength and $e$-$h$ asymmetry}. \textbf{a}, Spatial dependence of d$\mu$/d$n$ as a function of carrier density $n$ at $T = 1.6$ K (left panel) and the corresponding local twist angle $\theta_{12}$ (right panel). \textbf{b}, d$\mu$/d$n$ as a function of $\nu$ for select twist angles from the line cut in \textbf{a}. The correlated insulators (CI) at $\nu=\pm2$ and the sawtooth of d$\mu$/d$n$ are circled with black dashed lines. Curves are offset for clarity. \textbf{c}, The change in the chemical potential across the conduction band ($\Delta \mu_e$; blue), and the valence band ($\Delta \mu_h$; red) as a function of $\theta_{12}$. \textbf{d}, The gap of the correlated insulator $\Delta_{\pm2}$ at $\nu=\pm2$  (blue and red, respectively), as a function of $\theta_{12}$.
    \textbf{e}, The strength of the sawtooth (Methods) in d$\mu$/d$n$ for electron (hole) doping $\sigma_{e(h)}$ (blue and red, respectively), as a function of $\theta_{12}$.} 
    \label{fig:fig2}
\end{figure*}

\section*{Electronic structure}

We first describe the electronic characteristics relevant to our TTG device. Figure~\ref{fig:fig1}\textbf{b} shows the calculated spectral function along the path $K_3 \rightarrow K_2 \rightarrow K_1$ for $(\theta_{12}, \theta_{23}) =  (1.1^{\circ}, 3.5^{\circ})$ at zero displacement field (Methods; see Extended Data Figure 1 for calculations at other twist angle combinations and displacement fields). The colors represent the relative weight on each respective graphene layer (inset of Fig.~\ref{fig:fig1}\textbf{b}). The flat bands predominately originate from the strong hybridization between layers 1 and 2 due to their small interlayer twist. Layers 2 and 3 are more weakly coupled at low energies, resulting in a Dirac-like cone centered at $K_3$ predominantly associated with layer 3. It has a renormalized Fermi velocity $v^*_F < v^0_{F}$, where $v^0_{F}$ is the Fermi velocity of bare graphene. 
The result is an electronic system with coexisting light and heavy fermions where correlation effects are anticipated when the Fermi energy lies within the flat bands. 

We experimentally probe the electronic structure and resulting ground states by measuring the local inverse electronic compressibility d$\mu$/d$n$ with a high-resolution scanning SET (Methods).
Representative traces of d$\mu$/d$n$ and the corresponding chemical potential $\mu$ as a function of filling at a fixed position are shown in Fig.~\ref{fig:fig1}\textbf{c}. We identify the local twist angle $\theta_{12} = 1.10^\circ$ from the density $n$ of the two largest incompressible peaks, which are associated with full-filling ($\nu=\pm4$) of the moiré unit cell created by layers 1 and 2 (Methods). We estimate $\theta_{23} \approx 3.45^{\circ}$ based on the extracted Fermi velocity $v^*_F$ of the Landau levels from the Dirac-like cone in a magnetic field (Methods and Extended Data Figure 2). At intermediate fillings, we find a sharp incompressible state at $\nu=2$, indicating a correlated insulator with a gap $\Delta_2 = 1$~meV, as well as more slowly varying, asymmetric oscillations of d$\mu$/d$n$ between $0 < \nu < 4$. These oscillations resemble a sawtooth shape, and we adopt the ``sawtooth" terminology for brevity below. Both of the above features closely match ubiquitous signatures of strong correlations reported in other twisted graphene systems~\cite{cao2018correlated,sharpe2019emergent,yankowitz2019tuning,lu2019superconductors,yu_spin_2022,pierce2025tunable, zondiner2020cascade,yu2022correlated,saito2021isospin,park2021tunable,liu2022isospin}. Lastly, we note that while dispersive Dirac-like bands coexist with the flat bands, their filling is relatively small in our measurements. Consequently, the Dirac-like bands remain incompressible compared to the flat bands, and the SET signal is dominated by the flat bands~\cite{pierce2025tunable}.

\section*{Mapping twist-dependent signatures of interaction effects}

\begin{figure*}[t!]
    \renewcommand{\thefigure}{\arabic{figure}}
    \centering
    \includegraphics[width=1.99\columnwidth]{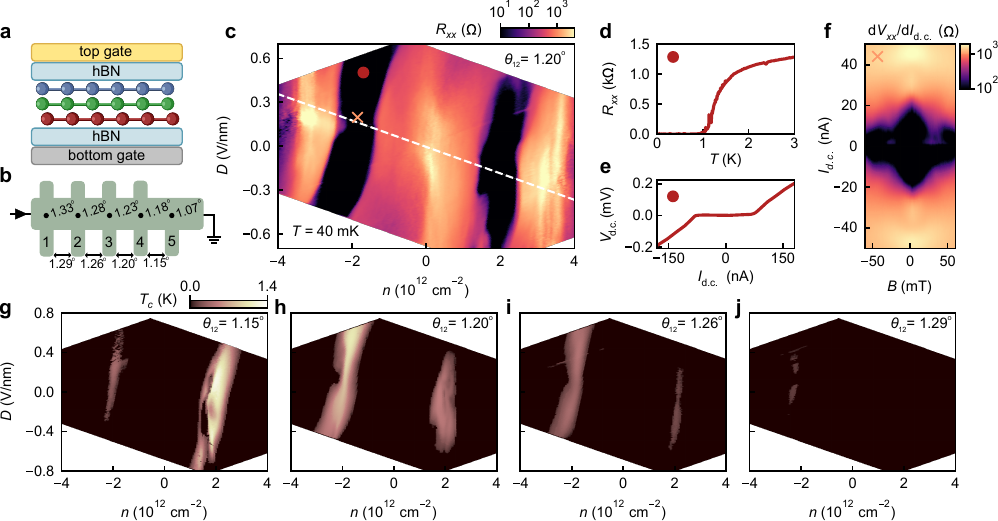}
    \caption{\textbf{Dual-gated transport and superconductivity}. \textbf{a}-\textbf{b}, Schematic of the dual-gated device structure (\textbf{a}) and its Hall bar geometry (\textbf{b}). Contacts 1-5 are labeled, along with the observed twist angles $\theta_{12}$ between contact pairs as measured in transport (bottom), and from local compressibility measurements in the Hall bar (black dots; see Extended Data Fig. 3). \textbf{c}, Longitudinal resistance $R_{xx}$ between contacts 3 and 4 as a function of $n$ and displacement field $D$. The white dashed line corresponds to where the top gate voltage $V_t = 0$, which most closely approximates the conditions of SET measurements. \textbf{d}, $R_{xx}$ as a function of $T$ for contacts 3-4 at $n = -1.69\times10^{12}$~cm$^{-2}$ and $D = 0.50$~V/nm. \textbf{e}, d.c.~voltage $V_{\textrm{d.c.}}$ as a function of d.c.~current $I_{\textrm{d.c.}}$ for contacts 3-4 at $T=40$~mK, $n = -1.69\times10^{12}$~cm$^{-2}$, and $D = 0.50$~V/nm. The data in \textbf{d} and \textbf{e} were taken at the location denoted by the red circle in \textbf{c}. \textbf{f}, $\mathrm{d}V_{xx}/\mathrm{d}I_{\textrm{d.c.}}$ as a function of $I_{\textrm{d.c.}}$ and perpendicular magnetic field $B$ for contacts 3-4 at $T=40$~mK, $n = -2.35\times10^{12}$~cm$^{-2}$, and $D = -0.26$~V/nm. Oscillations in the critical current as a function of $B$ are emblematic of phase coherent transport. hese data were taken at the location in \textbf{c} denoted by the orange x. \textbf{g}-\textbf{j}, Critical temperature $T_c$ (defined as 10\% of the normal state resistance; Methods) as a function of $n$ and $D$ for each contact pair.}
    \label{fig:fig3}
\end{figure*}

To characterize the twist angle dependence of the correlations in TTG, we map d$\mu$/d$n$ as a function of spatial position
(Fig.~\ref{fig:fig2}\textbf{a}; see Extended Data Figure 3 for additional line cuts). Along this line cut, $\theta_{12}$ varies smoothly from $\theta_{12} \approx 1^\circ$ to $\theta_{12} \approx 1.4^\circ$ over a span of about 4~$\mu$m. 
We observe correlation-driven gaps at $\nu=\pm2$ and/or a sawtooth in compressibility over a range of twist angles $1.03^\circ < \theta_{12} < 1.24^\circ$. We take the gap size and sawtooth strength as proxies for the strength of interactions. However, these quantities are indirect measures of many-body correlations; strong correlations may, for example, also persist in metallic (gapless) states.
This range is consistent with the theoretically predicted magic continuum of angles for TTG where single-particle bandwidths are minimized for our $\theta_{23}$~\cite{zhu2020twisted,yang2024multi,foo2024extended}.
The data reveal that the locations where these correlation features are maximized for hole and electron doping differ, indicating distinct effective magic angles for the valence and conduction bands (Fig.~\ref{fig:fig2}\textbf{b}).
While differences between the valence and conduction band behaviors are frequently observed in twisted graphene samples, our spatially resolved measurements reveal that both bands can host correlations of similar strength, yet the detailed conditions that optimize interaction effects within each band are different.

To quantify the $\theta_{12}$ dependence and the $e$-$h$ asymmetry, we extract three key parameters from the line cut in Fig.~\ref{fig:fig2}\textbf{a}. 
The first is the change in chemical potential $\Delta\mu_{e({h})}$ to fully fill the conduction (valence) band, as defined in Fig.~\ref{fig:fig1}\textbf{c}. The second is the gap size $\Delta_{\pm2}$ of the correlated insulators at $\nu=\pm2$ (inset of Fig.~\ref{fig:fig1}\textbf{c}; see Methods). Third, we quantify the strength of the sawtooth in inverse compressibility as $\sigma_{e(h)}$, the standard deviation of d$\mu$/d$n$ between $0.5 < |\nu| < 3.5$ for electron and hole doping, respectively. A small (large) $\sigma_{e(h)}$ indicates a weak/nonexistent (strong) sawtooth variation. When calculating $\sigma_{e(h)}$, a filter is applied to smooth the peaks in d$\mu$/d$n$ associated with the $\nu = \pm2$ correlated insulators, ensuring these effects can be evaluated independently (Methods and Extended Data Figure 4). 

We plot $\Delta\mu_{e(h)}$, $\Delta_{\pm2}$, and $\sigma_{e(h)}$ as a function of $\theta_{12}$ in Fig.~\ref{fig:fig2}\textbf{c}-\textbf{e}, respectively. Interaction strength, as quantified by both $\Delta_{\pm2}$ and $\sigma_{e(h)}$, is maximized at different $\theta_{12}$ for conduction and valence bands. The $\theta_{12}$ where $\Delta_{\pm2}$ and $\sigma_{e(h)}$ are maximal are not coincident with the minimum of $\Delta \mu_{e(h)}$, as might be expected. Note, however, that $\Delta \mu$ is not equivalent to the non-interacting bandwidth because interactions significantly modify the band structure as the bands are filled, enhancing the many-body interacting $\Delta \mu$ extracted here~\cite{choi2021interaction,xiao2025interacting, yu_spin_2022}. 
In addition, $\sigma_e$ and $\Delta_2$ are themselves maximized at different $\theta_{12}$. 
The fact that these different features do not evolve together as a function of twist angle shows that the detailed conditions required to produce each are distinct.

While $1$D spatial line cuts provide a concrete quantitative understanding of how the electronic character varies with $\theta_{12}$, a natural question is the extent to which other microscopic parameters, such as local variations in $\theta_{23}$ or strain, may influence the resulting electronic properties. To address this, we conduct local measurements of d$\mu$/d$n$ spanning a $1 \times 1$ $\mu$m$^2$ area where $\theta_{12}$ varies between $1.07^\circ$ and $1.22^\circ$ (Extended Data Figure 5), extracting $\Delta\mu_f$ (the sum of $\Delta\mu_e$ and $\Delta\mu_h$), $\Delta_2$, and $\sigma_{e(h)}$ at each pixel. The data in Extended Data Figure 5\textbf{h}-\textbf{k} show clear trends of each of these parameters as a function $\theta_{12}$, demonstrating that $\theta_{12}$ is the dominant determinant of electronic behavior, consistent with its direct relation to the flat bands. However, the data in Extended Data Figure 5\textbf{h}-\textbf{k} also reveal systematic spatial variations in each of these parameters, even as $\theta_{12}$ remains fixed, demonstrating that $\theta_{23}$ and/or local strain can play a secondary, but non-negligible role in modifying the electronic structure.
Although the specific microscopic origins of this spatial dependence are difficult to determine conclusively, we speculate that the observed variations are more likely to result from spatial variability in $\theta_{23}$ than by changes in local strain (Methods).

\section*{Superconductivity}

\textbf{\begin{figure*}[t!]
    \renewcommand{\thefigure}{\arabic{figure}}
    \centering
    \includegraphics[width=1.8\columnwidth]{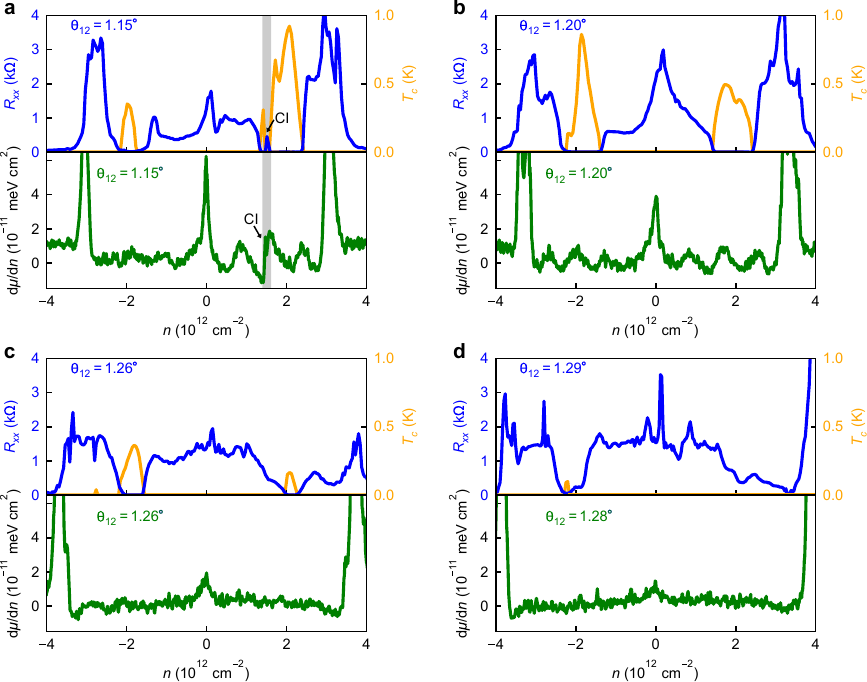}
    \caption{\textbf{Comparison of compressibility and transport at $V_t=0$}. \textbf{a}, Upper panel: line traces of longitudinal resistance $R_{xx}$ (blue) and critical temperature $T_c$ (orange) as a function of carrier density $n$ for the contact pair with twist angle $\theta_{12} = 1.15^\circ$. The line trace is taken with $V_t = 0$ (white dashed line in Fig.~\ref{fig:fig3}), which most closely approximates the conditions of SET measurements. Lower panel: line trace of the inverse compressibility d$\mu$/d$n$ as a function $n$ for a spatial location along line cut 3 in Extended Data Fig. 3 where $\theta_{12} = 1.15^\circ$. A correlated insulator (CI; labeled with arrows) is seen in both the compressibility and transport measurements at this twist angle. \textbf{b}-\textbf{d}, Same as \textbf{a}, but for the other contact pairs.
    }
    \label{fig:fig4}
\end{figure*}}

Having established the local twist angle dependence of thermodynamic correlation signatures, we next measure transport in the same area of the device for direct comparison. To do so, we patterned a large portion of the SET-characterized region into a Hall bar, using a deposited metal top gate as a hard etch mask (Extended Data Figure 6; Methods). 
The newly dual gated architecture allows independent control of the carrier density $n$ and displacement field $D$ (Fig.~\ref{fig:fig3}\textbf{a}). We oriented the long axis of the Hall bar along the local gradient in $\theta_{12}$ as measured by SET (Extended Data Figure 3). This enables us to correlate signatures in transport with angle-resolved features in electronic compressibility from the scanning SET. The twist angles $\theta_{12}$ determined from transport (Methods) quantitatively match those measured with the scanning SET in the vicinity of each contact pair (Fig.~\ref{fig:fig3}\textbf{b}, Extended Data Figure 3, indicating that the additional fabrication procedures did not meaningfully change the interlayer twists within the heterostructure.

Figure~\ref{fig:fig3}\textbf{c} shows the longitudinal resistance $R_{xx}$ as a function of $n$ and $D$ between contacts 3 and 4, whose $\theta_{12} = 1.20^\circ$. 
At temperature $T = 40$~mK, we observe two large superconducting pockets of zero resistance upon both electron and hole doping. The superconductivity is confirmed by standard temperature dependence, non-linear $I$-$V$ characteristics, and magnetic field measurements that reveal oscillations of the critical current, indicative of  phase coherent transport that arises from Josephson junction-like behavior likely stemming from disorder within the sample (Fig.~\ref{fig:fig3}\textbf{d}-\textbf{f}). 
The regions of superconductivity exhibit a slight tilt as $n$ and $D$ are varied (Fig.~\ref{fig:fig3}\textbf{c}), which likely originates from changes in the occupation of the Dirac-like band from the weakly coupled third layer as $D$ tunes the layer potentials. 
The transport features do not all evolve under variations in $n$ and $D$ with the characteristic S-shape (Extended Data Figure 7) that was previously seen when layer 3 becomes fully decoupled for $\theta_{23} \gtrsim 5^\circ$~\cite{hoke2024uncovering}. This indicates some hybridization of layer 3 with the other two, which is consistent with the observation of a renormalized Fermi velocity in the scanning SET measurements (Extended Data Figure 2). 

Measurements across different contact pairs demonstrate that superconducting regions persist as $\theta_{12}$ varies from $1.15^\circ$ to $1.29^\circ$ (Extended Data Figure 7). The superconductivity also exhibits $e$-$h$ asymmetry as a function of $\theta_{12}$, which is evident from the measured critical temperature $T_c$ (defined to be the temperature where $R_{xx}$ reaches 10\% of the normal state resistance; Methods) of the electron- and hope-doped superconducting phases as a function of $n$ and $D$ (Fig.~\ref{fig:fig3}\textbf{g}-\textbf{j}). In particular, the electron-doped superconductor exhibits its maximum critical temperature of 1.28~K at $\theta_{12} =1.15^\circ$ and the extent of the superconducting region and $T_c$ generally decrease with increasing $\theta_{12}$. In contrast, for hole-doping, superconductivity reaches its maximum critical temperature of 1.05~K in the contact pair with $\theta_{12} =1.20^\circ$. The extent of the hole-doped superconducting region and the $T_c$ generally decrease as $\theta_{12}$ is further increased or decreased. Thus, superconductivity is optimized at a larger $\theta_{12}$ for holes than for electrons, much like the signatures of interactions in local inverse compressibility. This $e$-$h$ asymmetry persists independent of $D$-field (Extended Data Figure 8), suggesting that the observed $e$-$h$ asymmetry is intrinsic to the flat bands and not dictated by any layer polarization induced by the $D$-field.

\section*{Comparing compressibility and transport}

\textbf{\begin{figure}[t!]
    \renewcommand{\thefigure}{\arabic{figure}}
    \centering
    \includegraphics[width=1\columnwidth]{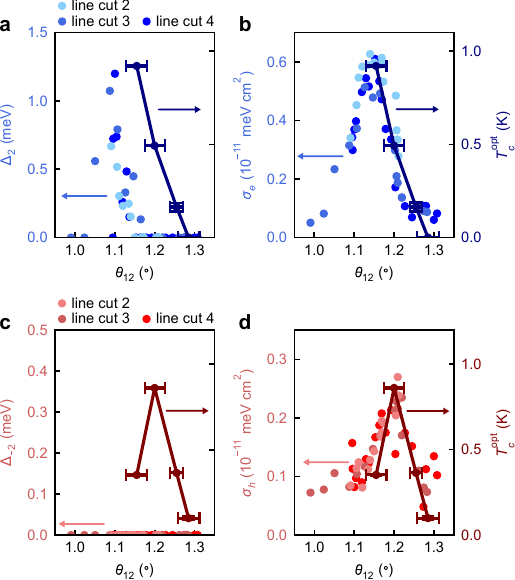}
    \caption{\textbf{Comparison of correlation signatures in compressibility and superconductivity in transport}. \textbf{a}, Correlated insulator gap $\Delta_2$ and critical temperature at optimal doping $T^{\mathrm{opt}}_c$ from the line traces for the electron-doped superconductor in Fig.~\ref{fig:fig4} as a function of $\theta_{12}$. \textbf{b}, Sawtooth compressibility strength $\sigma_e$ and critical temperature at optimal doping $T^{\mathrm{opt}}_c$ from the $V_t=0$ line traces for the electron-doped superconductor in Fig.~\ref{fig:fig4} as a function of $\theta_{12}$. \textbf{c}-\textbf{d}, Same as \textbf{a}-\textbf{b}, but for hole doping. For each panel distinct dot colors represent different line cuts (line cuts 2-4 in Extended Data Fig. 3) taken with the scanning SET that overlap with the Hall bar. The extraction of $\theta_{12}$ and $T^{\mathrm{opt}}_c$ for each contact pair are described in Methods. The horizontal error bars are set by the width of the superlattice peaks measured in transport, while the vertical error bars correspond to the standard deviation as measured via ``leave-out-three" Jackknife resampling (see Methods).
    }
    \label{fig:fig5}
\end{figure}}

We next directly compare the correlation signatures measured from compressibility with the superconductivity measured via transport. Due to the presence of only a bottom gate in SET experiments, $n$ and $D$ cannot be independently tuned. However, the SET tip can be approximated as an effective top gate with a voltage $V_t = 0$~\cite{foutty2023tunable}. Therefore, to most meaningfully and directly compare data taken with the SET and transport, we focus on transport data taken at a top gate voltage $V_t = 0$ (white dashed line in Fig.~\ref{fig:fig3}\textbf{c}). In Fig.~\ref{fig:fig4}\textbf{a}-\textbf{d} we plot line traces of $R_{xx}$ and $T_c$ as a function of $n$ at $V_t = 0$ for each contact pair, along with representative traces of d$\mu$/d$n$ at the same $\theta_{12}$. Aside from a thermodynamically gapped correlated insulator at $\nu=2$ for $\theta_{12} =1.15^\circ$, the data shows that, in general, superconductivity can exist without proximate correlated insulators.

To more quantitatively compare the different correlation signatures in compressibility and transport, we plot in Fig.~\ref{fig:fig5}\textbf{a}-\textbf{b} $\Delta_2$ and $\sigma_e$, respectively, as a function of $\theta_{12}$ (left axis) alongside the critical temperature $T^\mathrm{opt}_c$ at optimal doping (i.e. the maximum $T_c$) for each contact pair (right axis) from the line traces in Fig.~\ref{fig:fig4}. The corresponding results for $\Delta_{-2}$, $\sigma_h$, and $T^\mathrm{opt}_c$ of the hole-doped superconductor are shown in Fig.~\ref{fig:fig5}\textbf{c}-\textbf{d}. For the compressibility measurements, the dot colors denote different spatial line cuts taken with the SET microscope that overlap with the Hall bar (Extended Data Figure 3). Note that we do not include the data of Fig.~\ref{fig:fig2} here, which is adjacent to but outside the Hall bar (Extended Data Figure 3); a similar comparison with that data is shown in Extended Data Figure 9.

The twist angle dependencies of $\Delta_{\pm2}$ and $T^\mathrm{opt}_c$ do not identically overlap. 
With decreasing twist angle for electron-doping, we observe that superconductivity onsets at a higher twist angle than $\Delta_{2}$, forming a window of $\theta_{12}$ where $T^\mathrm{opt}_c > 0$ while $\Delta_{2} = 0$.
Notably, for the valence band, no thermodynamically gapped state is observed at $\nu=-2$ within the defined region of the Hall bar (Extended Data Figure 3), yet $T^\mathrm{opt}_c > 0$ for every contact pair. 
These observations are consistent with a picture in which the correlated insulator and superconducting phases are independent or competing~\cite{saito2020independent,stepanov2020untying,liu2021tuning}.

In contrast, comparing $\sigma_{e(h)}$ with $T^\mathrm{opt}_c$ uncovers a markedly closer correlation, with the relative strength of each signature displaying dependence on $\theta_{12}$ with near-identical scaling.
This correlation is present not only for the compressibility data shown here, but also that from the line cut in Fig.~\ref{fig:fig2} (Extended Data Figure 9) and the $1\times1$ $\mu$m$^2$ area of the sample (Extended Data Figure 10). Taken together, the close correlation between $\sigma_{e(h)}$ and $T^\mathrm{opt}_c$ suggests a deeper relationship between the two, which we discuss below. 

\section*{Conclusion}

The direct link between the strength of superconductivity and the sawtooth in compressibility raises interesting questions on the possible nature of the superconductivity. Recent experiments indicate that the sawtooth in compressibility in MATBG can be attributed to the redistribution of light and heavy charge carriers as the system is doped~\cite{xiao2025interacting, hu2024link,zhang2025heavy}. While itinerant conduction carriers are necessary for superconductivity in MATBG, it is possible heavy carriers and the interplay between the two charge sectors could also contribute~\cite{song2022magic}, in analogy to heavy fermion superconductors~\cite{steglich2016foundations}. Alternatively, it is possible that both signatures are independently maximized by the same underlying conditions. For example, a large density of states is favorable for superconductivity~\cite{tinkham2004introduction}, and it can also lead to Stoner-induced transitions in the occupation of distinct carriers. Our work motivates further theoretical efforts to examine why the sawtooth and superconductivity correlate so closely.

While our system has an additional coexisting dispersive Dirac-like band from the third layer, similar models of light and heavy fermions have been extended to systems hosting coexisting flat and dispersive Dirac-like bands~\cite{yu2023magic}.
In our experiment, superconductivity persists over a wide range of $D$, including where the filling of the Dirac-like band is negligible, suggesting that its carriers are not a prerequisite for superconductivity. However, we note that the extent of superconductivity is often markedly affected by the applied displacement field (Fig.~\ref{fig:fig3}\textbf{g}-\textbf{j}), suggesting states from the quasi-decoupled layer and/or changes in band structure can play a role. A detailed understanding of how superconductivity is modified by the filling of the Dirac-like band or the $D$-dependent hybridization between the coexisting flat and dispersive bands remains an open question and topic for future study.

We also report regions within the sample where both electron and hole-doped superconductors exist in the absence of proximate correlated insulators, consistent with the superconductors and correlated insulators being either competing or independent phases.  
The correlated insulators at $\nu=\pm2$ also appear to be more fragile, present only in a subset of the parameter space where we observe a sawtooth in d$\mu$/d$n$. This fragility is particularly pronounced for the correlated insulator at $\nu=-2$, which is only observed in a small region of the sample (Extended Data Figure 3). 
These observations align with proposals that the ordered phases at integer filling are inherently more delicate~\cite{kwan2021kekule,wagner2022global,liu2021nematic,parker2021strain,bi2019designing}, exhibiting greater sensitivity to strain, $\theta_{23}$, and/or disorder than the sawtooth in d$\mu$/d$n$, and, evidently, superconductivity.

The above insights were possible because we used local SET microscopy to first identify and characterize a region of a TTG sample with a range of slowly and smoothly varying $\theta_{12}$ of interest. Subsequently leveraging this knowledge to inform the placement of a Hall bar device then enabled a direct comparison between compressibility and transport without the confounding factors of sample-to-sample variations.
Our work demonstrates that $\theta_{12}$ is the key parameter determining strongly correlated behavior near $\theta_{12}/\theta_{23}\approx1/3$ and links distinct thermodynamic and transport signatures in TTG. An interesting direction for future work is to determine whether the observed correlation relations persist across other regions of the TTG magic continuum and in other twisted graphene multilayers. More generally, this ``local-to-global" approach is broadly applicable to elucidate the nature of correlated ground states in a variety of other van der Waals systems, which often host intertwined orders. Employing joint complementary experimental probes, as we have exemplified here, is a powerful route to disentangle the nature of correlated phases and discover relationships between them.


\section*{Acknowledgments}
We thank B. Andrei Bernevig, Xi Dai, Patrick Ledwith, Marc A. Kastner, Liqiao Xia, and Jiaojie Yan for helpful discussions. Scanning SET measurements were supported by the QSQM, an Energy Frontier Research Center funded by the U.S. Department of Energy (DOE), Office of Science, Basic Energy Sciences (BES), under Award \# DE-SC0021238. Transport studies were supported by National Science Foundation (Grant no. NSF-DMR-2237050). 
K.W. and T.T. acknowledge support from the JSPS KAKENHI (Grant Numbers 21H05233 and 23H02052), the CREST (JPMJCR24A5), JST and World Premier International Research Center Initiative (WPI), MEXT, Japan. J.C.H. acknowledges support from the Stanford Q-FARM Quantum Science and Engineering Fellowship. A.S. was supported by the US Department of Energy, Office of Science, Basic Energy Sciences, Materials Sciences and Engineering Division, under Contract DE-AC02-76SF00515. Measurement infrastructure was funded in part by the Gordon and Betty Moore Foundation’s EPiQS initiative through grant GBMF9460. Part of this work was performed at the Stanford Nano Shared Facilities (SNSF), supported by the National Science Foundation under award ECCS-2026822.

\section*{Author contributions}
J.C.H. and Y.L. fabricated the device. J.C.H. performed the scanning SET measurements. A.S., Y.L., and J.C.H. performed transport measurements. T.D. performed theoretical calculations. B.E.F supervised the project. K.W. and T.T provided hBN crystals. All authors participated in analysis and writing of the manuscript. 

\section*{Competing interests}
The authors declare no competing interest.

\newpage


\bibliographystyle{apsrev4-1}
\bibliography{references.bib}

@article{Andrei_2020,
  title={Graphene bilayers with a twist},
  author={Andrei, Eva Y and MacDonald, Allan H},
  journal={Nature Materials},
  volume={19},
  number={12},
  pages={1265--1275},
  year={2020},
  publisher={Nature Publishing Group UK London}
}

@article{Balents_2020,
  title={Superconductivity and strong correlations in moir{\'e} flat bands},
  author={Balents, Leon and Dean, Cory R and Efetov, Dmitri K and Young, Andrea F},
  journal={Nature Physics},
  volume={16},
  number={7},
  pages={725--733},
  year={2020},
  publisher={Nature Publishing Group UK London}
}

@article{Mak_2022,
  title={Semiconductor moir{\'e} materials},
  author={Mak, Kin Fai and Shan, Jie},
  journal={Nature Nanotechnology},
  volume={17},
  number={7},
  pages={686--695},
  year={2022},
  publisher={Nature Publishing Group UK London}
}

@article{bistritzer2011moire,
  title={Moir{\'e} bands in twisted double-layer graphene},
  author={Bistritzer, Rafi and MacDonald, Allan H},
  journal={Proceedings of the National Academy of Sciences},
  volume={108},
  number={30},
  pages={12233--12237},
  year={2011},
  publisher={National Academy of Sciences}
}

@article{cao2018correlated,
  title={Correlated insulator behaviour at half-filling in magic-angle graphene superlattices},
  author={Cao, Yuan and Fatemi, Valla and Demir, Ahmet and Fang, Shiang and Tomarken, Spencer L and Luo, Jason Y and Sanchez-Yamagishi, Javier D and Watanabe, Kenji and Taniguchi, Takashi and Kaxiras, Efthimios and others},
  journal={Nature},
  volume={556},
  number={7699},
  pages={80--84},
  year={2018},
  publisher={Nature Publishing Group UK London}
}

@article{cao2018unconventional,
  title={Unconventional superconductivity in magic-angle graphene superlattices},
  author={Cao, Yuan and Fatemi, Valla and Fang, Shiang and Watanabe, Kenji and Taniguchi, Takashi and Kaxiras, Efthimios and Jarillo-Herrero, Pablo},
  journal={Nature},
  volume={556},
  number={7699},
  pages={43--50},
  year={2018},
  publisher={Nature Publishing Group UK London}
}

@article{sharpe2019emergent,
  title={Emergent ferromagnetism near three-quarters filling in twisted bilayer graphene},
  author={Sharpe, Aaron L and Fox, Eli J and Barnard, Arthur W and Finney, Joe and Watanabe, Kenji and Taniguchi, Takashi and Kastner, MA and Goldhaber-Gordon, David},
  journal={Science},
  volume={365},
  number={6453},
  pages={605--608},
  year={2019},
  publisher={American Association for the Advancement of Science}
}

@article{yankowitz2019tuning,
  title={Tuning superconductivity in twisted bilayer graphene},
  author={Yankowitz, Matthew and Chen, Shaowen and Polshyn, Hryhoriy and Zhang, Yuxuan and Watanabe, Kenji and Taniguchi, Takashi and Graf, David and Young, Andrea F and Dean, Cory R},
  journal={Science},
  volume={363},
  number={6431},
  pages={1059--1064},
  year={2019},
  publisher={American Association for the Advancement of Science}
}

@article{lu2019superconductors,
  title={Superconductors, orbital magnets and correlated states in magic-angle bilayer graphene},
  author={Lu, Xiaobo and Stepanov, Petr and Yang, Wei and Xie, Ming and Aamir, Mohammed Ali and Das, Ipsita and Urgell, Carles and Watanabe, Kenji and Taniguchi, Takashi and Zhang, Guangyu and others},
  journal={Nature},
  volume={574},
  number={7780},
  pages={653--657},
  year={2019},
  publisher={Nature Publishing Group UK London}
}

@article{yu_spin_2022,
  title={Spin skyrmion gaps as signatures of strong-coupling insulators in magic-angle twisted bilayer graphene},
  author={Yu, Jiachen and Foutty, Benjamin A and Kwan, Yves H and Barber, Mark E and Watanabe, Kenji and Taniguchi, Takashi and Shen, Zhi-Xun and Parameswaran, Siddharth A and Feldman, Benjamin E},
  journal={Nature Communications},
  volume={14},
  number={1},
  pages={6679},
  year={2023},
  publisher={Nature Publishing Group UK London}
}

@article{zondiner2020cascade,
  title={Cascade of phase transitions and Dirac revivals in magic-angle graphene},
  author={Zondiner, Uri and Rozen, Asaf and Rodan-Legrain, Daniel and Cao, Yuan and Queiroz, Raquel and Taniguchi, Takashi and Watanabe, Kenji and Oreg, Yuval and von Oppen, Felix and Stern, Ady and others},
  journal={Nature},
  volume={582},
  number={7811},
  pages={203--208},
  year={2020},
  publisher={Nature Publishing Group UK London}
}

@article{yu2022correlated,
  title={Correlated Hofstadter spectrum and flavour phase diagram in magic-angle twisted bilayer graphene},
  author={Yu, Jiachen and Foutty, Benjamin A and Han, Zhaoyu and Barber, Mark E and Schattner, Yoni and Watanabe, Kenji and Taniguchi, Takashi and Phillips, Philip and Shen, Zhi-Xun and Kivelson, Steven A and others},
  journal={Nature Physics},
  volume={18},
  number={7},
  pages={825--831},
  year={2022},
  publisher={Nature Publishing Group UK London}
}

@article{saito2021isospin,
  title={Isospin Pomeranchuk effect in twisted bilayer graphene},
  author={Saito, Yu and Yang, Fangyuan and Ge, Jingyuan and Liu, Xiaoxue and Taniguchi, Takashi and Watanabe, Kenji and Li, JIA and Berg, Erez and Young, Andrea F},
  journal={Nature},
  volume={592},
  number={7853},
  pages={220--224},
  year={2021},
  publisher={Nature Publishing Group UK London}
}

@article{stepanov2020untying,
  title={Untying the insulating and superconducting orders in magic-angle graphene},
  author={Stepanov, Petr and Das, Ipsita and Lu, Xiaobo and Fahimniya, Ali and Watanabe, Kenji and Taniguchi, Takashi and Koppens, Frank HL and Lischner, Johannes and Levitov, Leonid and Efetov, Dmitri K},
  journal={Nature},
  volume={583},
  number={7816},
  pages={375--378},
  year={2020},
  publisher={Nature Publishing Group UK London}
}

@article{saito2020independent,
  title={Independent superconductors and correlated insulators in twisted bilayer graphene},
  author={Saito, Yu and Ge, Jingyuan and Watanabe, Kenji and Taniguchi, Takashi and Young, Andrea F},
  journal={Nature Physics},
  volume={16},
  number={9},
  pages={926--930},
  year={2020},
  publisher={Nature Publishing Group UK London}
}

@article{liu2021tuning,
  title={Tuning electron correlation in magic-angle twisted bilayer graphene using Coulomb screening},
  author={Liu, Xiaoxue and Wang, Zhi and Watanabe, Kenji and Taniguchi, Takashi and Vafek, Oskar and Li, JIA},
  journal={Science},
  volume={371},
  number={6535},
  pages={1261--1265},
  year={2021},
  publisher={American Association for the Advancement of Science}
}

@article{gao2024double,
  title={Double-edged Role of Interactions in Superconducting Twisted Bilayer Graphene},
  author={Gao, Xueshi and Jimeno-Pozo, Alejandro and Pantaleon, Pierre A and Codecido, Emilio and Sharifi, Daria L and Zhang, Zheneng and Liu, Youwei and Watanabe, Kenji and Taniguchi, Takashi and Bockrath, Marc W and others},
  journal={arXiv preprint arXiv:2412.01578},
  year={2024}
}

@article{barrier2024coulomb,
  title={Coulomb screening of superconductivity in magic-angle twisted bilayer graphene},
  author={Barrier, Julien and Peng, Liangtao and Xu, Shuigang and Fal'ko, VI and Watanabe, K and Tanigushi, T and Geim, AK and Adam, S and Berdyugin, Alexey I},
  journal={arXiv preprint arXiv:2412.01577},
  year={2024}
}

@article{wang2025independently,
  title={Independently Tunable Flat Bands and Correlations in a Graphene Double Moir{\'e} System},
  author={Wang, Yimeng and Zhu, Jihang and Burg, G William and Swain, Anand and Watanabe, Kenji and Taniguchi, Takashi and Zheng, Yuebing and MacDonald, Allan H and Tutuc, Emanuel},
  journal={Physical Review Letters},
  volume={134},
  number={9},
  pages={096204},
  year={2025},
  publisher={APS}
}

@article{tanaka2025superfluid,
  title={Superfluid stiffness of magic-angle twisted bilayer graphene},
  author={Tanaka, Miuko and Wang, Joel {\^I}-j and Dinh, Thao H and Rodan-Legrain, Daniel and Zaman, Sameia and Hays, Max and Almanakly, Aziza and Kannan, Bharath and Kim, David K and Niedzielski, Bethany M and others},
  journal={Nature},
  volume={638},
  number={8049},
  pages={99--105},
  year={2025},
  publisher={Nature Publishing Group UK London}
}

@article{cao2021nematicity,
  title={Nematicity and competing orders in superconducting magic-angle graphene},
  author={Cao, Yuan and Rodan-Legrain, Daniel and Park, Jeong Min and Yuan, Noah FQ and Watanabe, Kenji and Taniguchi, Takashi and Fernandes, Rafael M and Fu, Liang and Jarillo-Herrero, Pablo},
  journal={science},
  volume={372},
  number={6539},
  pages={264--271},
  year={2021},
  publisher={American Association for the Advancement of Science}
}

@article{arora2020superconductivity,
  title={Superconductivity in metallic twisted bilayer graphene stabilized by WSe2},
  author={Arora, Harpreet Singh and Polski, Robert and Zhang, Yiran and Thomson, Alex and Choi, Youngjoon and Kim, Hyunjin and Lin, Zhong and Wilson, Ilham Zaky and Xu, Xiaodong and Chu, Jiun-Haw and others},
  journal={Nature},
  volume={583},
  number={7816},
  pages={379--384},
  year={2020},
  publisher={Nature Publishing Group UK London}
}

@article{choi2021interaction,
  title={Interaction-driven band flattening and correlated phases in twisted bilayer graphene},
  author={Choi, Youngjoon and Kim, Hyunjin and Lewandowski, Cyprian and Peng, Yang and Thomson, Alex and Polski, Robert and Zhang, Yiran and Watanabe, Kenji and Taniguchi, Takashi and Alicea, Jason and others},
  journal={Nature Physics},
  volume={17},
  number={12},
  pages={1375--1381},
  year={2021},
  publisher={Nature Publishing Group UK London}
}

@article{xiao2025interacting,
  title={The Interacting Energy Bands of Magic Angle Twisted Bilayer Graphene Revealed by the Quantum Twisting Microscope},
  author={Xiao, J and Inbar, A and Birkbeck, J and Gershon, N and Zamir, Y and Taniguchi, T and Watanabe, K and Berg, E and Ilani, S},
  journal={arXiv preprint arXiv:2506.20738},
  year={2025}
}

@article{hu2024link,
  title={Link between cascade transitions and correlated Chern insulators in magic-angle twisted bilayer graphene},
  author={Hu, Qianying and Liang, Shu and Li, Xinheng and Shi, Hao and Dai, Xi and Xu, Yang},
  journal={arXiv preprint arXiv:2406.08734},
  year={2024}
}

@article{zhang2025heavy,
  title={Heavy fermions, mass renormalization and local moments in magic-angle twisted bilayer graphene via planar tunneling spectroscopy},
  author={Zhang, Zhenyuan and Wu, Shuang and C{\u{a}}lug{\u{a}}ru, Dumitru and Hu, Haoyu and Taniguchi, Takashi and Wanatabe, Kenji and Bernevig, Andrei B and Andrei, Eva Y},
  journal={arXiv preprint arXiv:2503.17875},
  year={2025}
}

@article{song2022magic,
  title={Magic-angle twisted bilayer graphene as a topological heavy fermion problem},
  author={Song, Zhi-Da and Bernevig, B Andrei},
  journal={Physical review letters},
  volume={129},
  number={4},
  pages={047601},
  year={2022},
  publisher={APS}
}

@article{yu2023magic,
  title={Magic-angle twisted symmetric trilayer graphene as a topological heavy-fermion problem},
  author={Yu, Jiabin and Xie, Ming and Bernevig, B Andrei and Das Sarma, Sankar},
  journal={Physical Review B},
  volume={108},
  number={3},
  pages={035129},
  year={2023},
  publisher={APS}
}

@article{kwan2021kekule,
  title={Kekul{\'e} spiral order at all nonzero integer fillings in twisted bilayer graphene},
  author={Kwan, Yves H and Wagner, Glenn and Soejima, Tomohiro and Zaletel, Michael P and Simon, Steven H and Parameswaran, Siddharth A and Bultinck, Nick},
  journal={Physical Review X},
  volume={11},
  number={4},
  pages={041063},
  year={2021},
  publisher={APS}
}

@article{wagner2022global,
  title={Global phase diagram of the normal state of twisted bilayer graphene},
  author={Wagner, Glenn and Kwan, Yves H and Bultinck, Nick and Simon, Steven H and Parameswaran, SA},
  journal={Physical review letters},
  volume={128},
  number={15},
  pages={156401},
  year={2022},
  publisher={APS}
}

@article{liu2021nematic,
  title={Nematic topological semimetal and insulator in magic-angle bilayer graphene at charge neutrality},
  author={Liu, Shang and Khalaf, Eslam and Lee, Jong Yeon and Vishwanath, Ashvin},
  journal={Physical Review Research},
  volume={3},
  number={1},
  pages={013033},
  year={2021},
  publisher={APS}
}

@article{parker2021strain,
  title={Strain-induced quantum phase transitions in magic-angle graphene},
  author={Parker, Daniel E and Soejima, Tomohiro and Hauschild, Johannes and Zaletel, Michael P and Bultinck, Nick},
  journal={Physical review letters},
  volume={127},
  number={2},
  pages={027601},
  year={2021},
  publisher={APS}
}

@article{bi2019designing,
  title={Designing flat bands by strain},
  author={Bi, Zhen and Yuan, Noah FQ and Fu, Liang},
  journal={Physical Review B},
  volume={100},
  number={3},
  pages={035448},
  year={2019},
  publisher={APS}
}

@article{zhang2021correlated,
  title={Correlated insulating states and transport signature of superconductivity in twisted trilayer graphene superlattices},
  author={Zhang, Xi and Tsai, Kan-Ting and Zhu, Ziyan and Ren, Wei and Luo, Yujie and Carr, Stephen and Luskin, Mitchell and Kaxiras, Efthimios and Wang, Ke},
  journal={Physical review letters},
  volume={127},
  number={16},
  pages={166802},
  year={2021},
  publisher={APS}
}

@article{park2021tunable,
  title={Tunable strongly coupled superconductivity in magic-angle twisted trilayer graphene},
  author={Park, Jeong Min and Cao, Yuan and Watanabe, Kenji and Taniguchi, Takashi and Jarillo-Herrero, Pablo},
  journal={Nature},
  volume={590},
  number={7845},
  pages={249--255},
  year={2021},
  publisher={Nature Publishing Group UK London}
}

@article{hao2021electric,
  title={Electric field--tunable superconductivity in alternating-twist magic-angle trilayer graphene},
  author={Hao, Zeyu and Zimmerman, AM and Ledwith, Patrick and Khalaf, Eslam and Najafabadi, Danial Haie and Watanabe, Kenji and Taniguchi, Takashi and Vishwanath, Ashvin and Kim, Philip},
  journal={Science},
  volume={371},
  number={6534},
  pages={1133--1138},
  year={2021},
  publisher={American Association for the Advancement of Science}
}

@article{pierce2025tunable,
  title={Tunable interplay between light and heavy electrons in twisted trilayer graphene},
  author={Pierce, Andrew T and Xie, Yonglong and Park, Jeong Min and Cai, Zhuozhen and Watanabe, Kenji and Taniguchi, Takashi and Jarillo-Herrero, Pablo and Yacoby, Amir},
  journal={Nature Physics},
  pages={1--6},
  year={2025},
  publisher={Nature Publishing Group UK London}
}

@article{liu2022isospin,
  title={Isospin order in superconducting magic-angle twisted trilayer graphene},
  author={Liu, Xiaoxue and Zhang, Naiyuan James and Watanabe, K and Taniguchi, T and Li, JIA},
  journal={Nature Physics},
  volume={18},
  number={5},
  pages={522--527},
  year={2022},
  publisher={Nature Publishing Group UK London}
}

@article{cao2021pauli,
  title={Pauli-limit violation and re-entrant superconductivity in moir{\'e} graphene},
  author={Cao, Yuan and Park, Jeong Min and Watanabe, Kenji and Taniguchi, Takashi and Jarillo-Herrero, Pablo},
  journal={Nature},
  volume={595},
  number={7868},
  pages={526--531},
  year={2021},
  publisher={Nature Publishing Group UK London}
}

@article{banerjee2025superfluid,
  title={Superfluid stiffness of twisted trilayer graphene superconductors},
  author={Banerjee, Abhishek and Hao, Zeyu and Kreidel, Mary and Ledwith, Patrick and Phinney, Isabelle and Park, Jeong Min and Zimmerman, Andrew and Wesson, Marie E and Watanabe, Kenji and Taniguchi, Takashi and others},
  journal={Nature},
  volume={638},
  number={8049},
  pages={93--98},
  year={2025},
  publisher={Nature Publishing Group UK London}
}

@article{hoke2024imaging,
  title={Imaging supermoire relaxation and conductive domain walls in helical trilayer graphene},
  author={Hoke, Jesse C and Li, Yifan and Hu, Yuwen and May-Mann, Julian and Watanabe, Kenji and Taniguchi, Takashi and Devakul, Trithep and Feldman, Benjamin E},
  journal={arXiv preprint arXiv:2410.16269},
  year={2024}
}

@article{xia2025topological,
  title={Topological bands and correlated states in helical trilayer graphene},
  author={Xia, Li-Qiao and de la Barrera, Sergio C and Uri, Aviram and Sharpe, Aaron and Kwan, Yves H and Zhu, Ziyan and Watanabe, Kenji and Taniguchi, Takashi and Goldhaber-Gordon, David and Fu, Liang and others},
  journal={Nature Physics},
  volume={21},
  number={2},
  pages={239--244},
  year={2025},
  publisher={Nature Publishing Group UK London}
}

@article{uri2023superconductivity,
  title={Superconductivity and strong interactions in a tunable moir{\'e} quasicrystal},
  author={Uri, Aviram and de la Barrera, Sergio C and Randeria, Mallika T and Rodan-Legrain, Daniel and Devakul, Trithep and Crowley, Philip JD and Paul, Nisarga and Watanabe, Kenji and Taniguchi, Takashi and Lifshitz, Ron and others},
  journal={Nature},
  volume={620},
  number={7975},
  pages={762--767},
  year={2023},
  publisher={Nature Publishing Group UK London}
}

@article{zhu2020twisted,
  title={Twisted trilayer graphene: A precisely tunable platform for correlated electrons},
  author={Zhu, Ziyan and Carr, Stephen and Massatt, Daniel and Luskin, Mitchell and Kaxiras, Efthimios},
  journal={Physical review letters},
  volume={125},
  number={11},
  pages={116404},
  year={2020},
  publisher={APS}
}

@article{yang2024multi,
  title={Multi-moir{\'e} trilayer graphene: lattice relaxation, electronic structure, and magic angles},
  author={Yang, Charles and May-Mann, Julian and Zhu, Ziyan and Devakul, Trithep},
  journal={Physical Review B},
  volume={110},
  number={11},
  pages={115434},
  year={2024},
  publisher={APS}
}

@article{foo2024extended,
  title={Extended magic phase in twisted graphene multilayers},
  author={Foo, DCW and Zhan, Z and Al Ezzi, Mohammed M and Peng, L and Adam, S and Guinea, Francisco},
  journal={Physical Review Research},
  volume={6},
  number={1},
  pages={013165},
  year={2024},
  publisher={APS}
}

@article{turkel2022orderly,
  title={Orderly disorder in magic-angle twisted trilayer graphene},
  author={Turkel, Simon and Swann, Joshua and Zhu, Ziyan and Christos, Maine and Watanabe, K and Taniguchi, T and Sachdev, Subir and Scheurer, Mathias S and Kaxiras, Efthimios and Dean, Cory R and others},
  journal={Science},
  volume={376},
  number={6589},
  pages={193--199},
  year={2022},
  publisher={American Association for the Advancement of Science}
}

@article{hoke2024uncovering,
  title={Uncovering the spin ordering in magic-angle graphene via edge state equilibration},
  author={Hoke, Jesse C and Li, Yifan and May-Mann, Julian and Watanabe, Kenji and Taniguchi, Takashi and Bradlyn, Barry and Hughes, Taylor L and Feldman, Benjamin E},
  journal={Nature communications},
  volume={15},
  number={1},
  pages={4321},
  year={2024},
  publisher={Nature Publishing Group UK London}
}

@article{xie2025strong,
  title={Strong interactions and isospin symmetry breaking in a supermoir{\'e} lattice},
  author={Xie, Yonglong and Pierce, Andrew T and Park, Jeong Min and Parker, Daniel E and Wang, Jie and Ledwith, Patrick and Cai, Zhuozhen and Watanabe, Kenji and Taniguchi, Takashi and Khalaf, Eslam and others},
  journal={Science},
  pages={eadl2544},
  year={2025},
  publisher={American Association for the Advancement of Science}
}

@article{kim2022evidence,
  title={Evidence for unconventional superconductivity in twisted trilayer graphene},
  author={Kim, Hyunjin and Choi, Youngjoon and Lewandowski, Cyprian and Thomson, Alex and Zhang, Yiran and Polski, Robert and Watanabe, Kenji and Taniguchi, Takashi and Alicea, Jason and Nadj-Perge, Stevan},
  journal={Nature},
  volume={606},
  number={7914},
  pages={494--500},
  year={2022},
  publisher={Nature Publishing Group UK London}
}

@article{kim2023imaging,
  title={Imaging inter-valley coherent order in magic-angle twisted trilayer graphene},
  author={Kim, Hyunjin and Choi, Youngjoon and Lantagne-Hurtubise, {\'E}tienne and Lewandowski, Cyprian and Thomson, Alex and Kong, Lingyuan and Zhou, Haoxin and Baum, Eli and Zhang, Yiran and Holleis, Ludwig and others},
  journal={Nature},
  volume={623},
  number={7989},
  pages={942--948},
  year={2023},
  publisher={Nature Publishing Group UK London}
}

@article{shen2023dirac,
  title={Dirac spectroscopy of strongly correlated phases in twisted trilayer graphene},
  author={Shen, Cheng and Ledwith, Patrick J and Watanabe, Kenji and Taniguchi, Takashi and Khalaf, Eslam and Vishwanath, Ashvin and Efetov, Dmitri K},
  journal={Nature Materials},
  volume={22},
  number={3},
  pages={316--321},
  year={2023},
  publisher={Nature Publishing Group UK London}
}

@article{chen2020tunable,
  title={Tunable correlated Chern insulator and ferromagnetism in a moir{\'e} superlattice},
  author={Chen, Guorui and Sharpe, Aaron L and Fox, Eli J and Zhang, Ya-Hui and Wang, Shaoxin and Jiang, Lili and Lyu, Bosai and Li, Hongyuan and Watanabe, Kenji and Taniguchi, Takashi and others},
  journal={Nature},
  volume={579},
  number={7797},
  pages={56--61},
  year={2020},
  publisher={Nature Publishing Group UK London}
}

@article{zhang2022promotion,
  title={Promotion of superconductivity in magic-angle graphene multilayers},
  author={Zhang, Yiran and Polski, Robert and Lewandowski, Cyprian and Thomson, Alex and Peng, Yang and Choi, Youngjoon and Kim, Hyunjin and Watanabe, Kenji and Taniguchi, Takashi and Alicea, Jason and others},
  journal={Science},
  volume={377},
  number={6614},
  pages={1538--1543},
  year={2022},
  publisher={American Association for the Advancement of Science}
}

@article{park2022robust,
  title={Robust superconductivity in magic-angle multilayer graphene family},
  author={Park, Jeong Min and Cao, Yuan and Xia, Li-Qiao and Sun, Shuwen and Watanabe, Kenji and Taniguchi, Takashi and Jarillo-Herrero, Pablo},
  journal={Nature Materials},
  volume={21},
  number={8},
  pages={877--883},
  year={2022},
  publisher={Nature Publishing Group UK London}
}

@book{tinkham2004introduction,
  title={Introduction to superconductivity},
  author={Tinkham, Michael},
  year={2004},
  publisher={Courier Corporation}
}

@article{steglich2016foundations,
  title={Foundations of heavy-fermion superconductivity: lattice Kondo effect and Mott physics},
  author={Steglich, Frank and Wirth, Steffen},
  journal={Reports on Progress in Physics},
  volume={79},
  number={8},
  pages={084502},
  year={2016},
  publisher={IOP Publishing}
}

@article{popov2023magic,
  title={Magic angle butterfly in twisted trilayer graphene},
  author={Popov, Fedor K and Tarnopolsky, Grigory},
  journal={Physical Review Research},
  volume={5},
  number={4},
  pages={043079},
  year={2023},
  publisher={APS}
}

@article{foutty2023tunable,
  title={Tunable spin and valley excitations of correlated insulators in $\Gamma$-valley moir{\'e} bands},
  author={Foutty, Benjamin A and Yu, Jiachen and Devakul, Trithep and Kometter, Carlos R and Zhang, Yang and Watanabe, Kenji and Taniguchi, Takashi and Fu, Liang and Feldman, Benjamin E},
  journal={Nature Materials},
  volume={22},
  number={6},
  pages={731--736},
  year={2023},
  publisher={Nature Publishing Group UK London}
}

\newpage


\section*{Methods}

\subsection*{Spectral function calculations}
In this section, we describe our method for calculating the electronic structure at incommensurate twist angles.
We use the method described in Ref.~\cite{uri2023superconductivity}, which we briefly review here.
An eigenstate is expressed as
\begin{equation}
\begin{split}
|\psi_{\bm{k},n}\rangle
&= \sum_{\bm{G}_2,\bm{G}_3,\alpha}\phi_{1,\bm{k},\alpha}^{(n)}|1,\bm{k}+\bm{G}_2+\bm{G}_3,\alpha\rangle\\
&\;\;\;\;\; + \sum_{\bm{G}_1,\bm{G}_3,\alpha}\phi_{2,\bm{k},\alpha}^{(n)}|2,\bm{k}+\bm{G}_1+\bm{G}_3,\alpha\rangle\\
&\;\;\;\;\; + \sum_{\bm{G}_1,\bm{G}_2,\alpha}\phi_{3,\bm{k},\alpha}^{(n)}|3,\bm{k}+\bm{G}_1+\bm{G}_2,\alpha\rangle\\
\end{split}
\end{equation}
where $|\ell,\bm{k},\alpha\rangle$ is the graphene Bloch state on layer $\ell$ at momentum $\bm{k}$ (defined modulo $\bm{G}_{\ell}$), $n$ is the eigenstate index, $\alpha=A,B$ is the sublattice (spin degeneracy is assumed throughout), $\bm{G}_{\ell}$ is the atomic reciprocal lattice vectors for layer $\ell$.
Relative to an unrotated graphene reciprocal lattice vector $\bm{G}_0$, 
$\bm{G}_{\ell}=\bm{R}(\theta_{\ell})\bm{G}_0$, where $\theta_{\ell}$ is the twist angle of layer $\ell$.

The $\phi^{(n)}_{\ell,\bm{k},\alpha}$ are complex wavefunction coefficients obtained by diagonalizing the Hamiltonian matrix, which consists of an intralayer and interlayer part. 
For the intralayer part, we take the Dirac dispersion.  
The non-zero matrix elements are given by
\begin{equation}
\begin{split}
\langle \ell, \bm{k}, \alpha|&H_{\ell}|\ell,\bm{k},\alpha^\prime\rangle=\\
&
v_F\begin{pmatrix}
U_\ell & (k_x - i k_y-[K_{\ell,x}-iK_{\ell,y}])e^{i\theta_{\ell}} \\
\text{c.c.} & U_\ell \\
\end{pmatrix}_{\alpha\alpha^\prime}
\end{split}
\end{equation}
where $\bm{K}_{\ell}$ is the graphene $K$-points of layer $\ell$.
The layer-dependent potential $U_{\ell}$ is used to model the effect of a displacement field, $(U_1,U_2,U_3)=(-u_D,0,u_D)$.

The interlayer tunneling term is given by
\begin{equation}
\begin{split}
\langle \ell,\bm{k}+\bm{G}_{123},\alpha|&H_{\ell\ell^\prime}|\ell^\prime,\bm{k}+\bm{G}_{123},\alpha^\prime\rangle =\\
&t(\bm{k}+\bm{G}_{123})e^{i\bm{G}_{\ell}\bm{\tau}^{\ell}_{\alpha}-i\bm{G}_{\ell^\prime}\cdot\bm{\tau}^{\ell}_{\alpha^\prime}}
\begin{pmatrix}
\kappa & 1\\
1 & \kappa
\end{pmatrix}_{\alpha\alpha^\prime}
\end{split}
\end{equation}
where $\bm{G}_{123}\equiv\bm{G}_1+\bm{G}_2+\bm{G}_3$ is the sum of any three reciprocal lattice vectors, and $\tau^{\ell}_{\alpha}$ are the positions of the sublattice $\alpha$ atoms in the graphene unit cell of layer $\ell$, and $\kappa$ is the chiral ratio suppressing $AA$ and $BB$ tunneling.
In this expression, we have used the momentum space periodicity $|\ell,\bm{k}+\bm{G}_\ell,\alpha\rangle\equiv |\ell,\bm{k},\alpha\rangle$.

For a given $\bm{k}$, the Hamiltonian matrix is constructed by considering all $\bm{G}_{123}$ up to a cutoff $|\bm{G}_{\ell}|<7.1K$.  
We take a linear approximation to the tunneling amplitude, $t(\bm{k})\approx t_0 + t_1(|\bm{k}|-K)$, where $K$ is the magnitude of the $K$ point momentum, for $0<|\bm{k}|<1.4K$ and zero otherwise --- for our choice of cutoff, this is equivalent to the first harmonic approximation of the Bistritzer-MacDonald model.
The matrix is then diagonalized to obtain the list of eigenenergies $\varepsilon_{n\bm{k}}$ and eigenvectors $\phi^{(n)}_{\ell,\bm{k},\alpha}$.
The layer-resolved spectral weight is obtained by $w_{n\bm{k}\ell}=\sum_{\alpha}|\phi^{(n)}_{\ell\bm{k}\alpha}(\bm{0},\bm{0})|^2$.
In the spectral functions shown in Fig.~\ref{fig:fig1} and Extended Data Figure 1, the energy $\varepsilon_{n\bm{k}}$ is plotted along a path in $\bm{k}$, with the width of the line proportional to the total spectral weight $\sum_{\ell}w_{n\bm{k}\ell}$ and the color indicating the direction of the vector $(w_{n\bm{k}1},w_{n\bm{k}2},w_{n\bm{k}3})$ as indicated in the color triangle. 
The parameters used are $v_F=0.88\times10^6$ m/s, $t_0=0.11$ eV, $t_1=-0.227$ eV\AA, and $\kappa=0.7$.
The calculations shown in Fig.~\ref{fig:fig1} and Extended Data Figure 1\textbf{a}-\textbf{f} are computed with $u_D=0$.  
Further displacement field dependence is shown in Extended Data Figure 1\textbf{g}-\textbf{i}.

\subsection*{Device fabrication}
The TTG heterostructure studied was fabricated using standard dry transfer techniques with poly (bisphenol A carbonate)/polydimethylsiloxane (PC/PDMS) transfer slides. A monolayer graphene flake was cut into 3 pieces with a conductive atomic force microscope (AFM) tip in contact mode. An exfoliated hBN flake (42 nm thick) was then used to sequentially pick up each section at the desired twist angle before placing it on top of a prefabricated stack of few-layer graphite and hBN (33 nm thick), which had been previously vacuum annealed at $400^\circ$ C for $8$ hours to ensure cleanliness of the surface. The full stack was then patterned with standard electron beam lithography techniques followed by etching and metallization to form edge contacts (Extended Data Figure 6\textbf{d}). After measurements in the SET microscope, the device was further patterned with a Cr/Au top metallic gate, etched into a Hall bar geometry, and patterned with additional edge contacts (Extended Data Figure 6\textbf{e}-\textbf{f}).

We note that although the addition of a metallic top gate introduces a potential new screening layer to the sample, the top gate sits atop a $d = 42$ nm thick hBN dielectric layer, which is significantly greater than the $\lambda_{12} \sim 10 - 15$ nm moiré wavelength associated with layers 1 and 2. Thus we expect the top gate to have negligible effects related to Coulomb screening, which only becomes significant when $d < \lambda_{12}$~\cite{saito2020independent,stepanov2020untying}.

\subsection*{Scanning SET measurements}
The SET tip was fabricated by evaporating aluminum onto a pulled quartz rod and has an estimated diameter of $100$~nm at its apex. The tip is brought near the sample surface and scanning SET measurements were performed in a Unisoku USM 1300 scanning probe microscope with a microscope head customized for scanning SET operation. A 5~mV a.c.~excitation at 823~Hz was applied to the sample, while a 20 mV~a.c.~excitation at 911.7~Hz was applied to the back gate. We then measure the inverse compressibility d$\mu$/d$n$ $\propto I_\mathrm{BG}/I_\mathrm{2D}$, where $I_\mathrm{BG}$ and $I_\mathrm{2D}$ are the measured SET current demodulated at the back gate and sample frequencies, respectively, using standard lock-in techniques. A d.c.~offset voltage $V_\mathrm{2D}$ is applied to the sample to maintain maximum sensitivity of the SET and minimize tip-induced doping. SET measurements were taken at $T = 1.6$~K or 330~mK, as specified in the figure captions.

\subsection*{Transport measurements}
Transport measurements were performed in a Leiden Cryogenics CF-900 dilution refrigerator using a custom probe. 
The measurement lines are equipped with electronic filtering thermally anchored at the mixing chamber stage to obtain a low electron temperature and reduce high-frequency noise. 
There are two stages of filtering. First, the wires are passed through a cured mixture of epoxy and bronze powder to filter GHz frequencies.
Subsequently, low-pass RC filters mounted on sapphire plates attenuate MHz frequencies. 
Samples were mounted using a Kyocera custom 32-contact ceramic leadless chip carrier (drawing PB-44567-Mod with no nickel sticking layer under gold, to reduce spurious magnetic effects).
Stanford Research Systems SR830 lock-in amplifiers with NF Corporation LI-75A voltage preamplifiers were used to perform four-terminal resistance measurements.
A 1~G$\Omega$ bias resistor was used to apply an a.c. bias current of up to 5~nA RMS at low frequencies.
Keithley 2400 source-measure units were used to apply voltages to the gates.

\subsection*{Determination of $\theta_{12}$}
The conversion from applied voltages to carrier density $n$ in the SET experiment is determined by the geometric capacitance between the graphite bottom gate and the TTG sample: $n = C_{\rm{b}}(V_{\rm{b}}-V_{\rm{2D}})/e$, where $V_{\rm{b}}-V_{\rm{2D}}$ is the difference in the voltages applied to the bottom gate and the sample, and $e$ is the electron charge. The capacitance $C_{\rm{b}}$ between the bottom gate and sample is determined from the slopes of quantum Hall features emanating from charge neutrality in Landau fan measurements, which are quantized according to fundamental constants. This extracted capacitance is consistent with a parallel plate capacitor where the separation between plates is equal to the hBN thickness measured by atomic force microscopy. 

In transport measurements, the dual gated geometry of the device allows for independent control of $n$ and the perpendicular electric displacement field $D$, where $n = C_{t}V_{t}/e + C_{b}V_{b}/e$ and $D =(C_{t}V_{t}/e - C_{b}V_{b}/e)/(2\epsilon_{0})$. Here, $V_{b}$ ($V_{t}$) is the voltage applied to the bottom (top) gate, $C_{b}$ ($C_{t}$) is the bottom (top) gate capacitance, and $\epsilon_{0}$ is the vacuum permittivity. The capacitance $C_{\rm{b}}$ ($C_{\rm{t}}$) between the bottom (top) gate and sample is again determined from the slopes of quantum Hall features emanating from charge neutrality in Landau fan measurements, which are quantized according to fundamental constants. The extracted $C_{\rm{b}}$ from transport experiments match with the extracted $C_{\rm{b}}$ from SET experiments.

In both SET and transport measurements, the superlattice carrier density $n_{s}$, associated with the twist angle between layers 1 and 2, is determined by fitting the $\nu=\pm4$ states in a finite magnetic field $B$ to the Str\v{e}da formula $\frac{\partial n}{\partial B} = \pm2 \, \frac{e}{h}$, where the $\pm2$ comes from the filling of the lowest Dirac-like Landau level, and extrapolating to $B=0$. In scanning SET measurements at $B = 0$, we instead take $n_s$ to be $n_s = (n(\nu=4) - n(\nu=-4))/2$, where $n(\nu=\pm4)$ is the carrier density of the peak of the $\nu=\pm4$ incompressible states. The interlayer twist angle $\theta_{12}$ is then found by $n_{s} = 4/A \approx 8\theta^{2}_{12}/\sqrt{3}a^{2}$. Here, $A$ is the moiré lattice unit cell area associated with the moiré superlattice of layers 1 and 2, and $a = 0.246$ nm is the graphene lattice constant. We find empirically that the two methods to determine $\theta_{12}$ give results with negligible differences. 

\subsection*{Hofstadter spectrum measurement}
The procedure to measure the correlated Hofstadter spectrum from inverse compressibility experiments is discussed in detail in the Supplement of Ref.~\cite{yu2022correlated}. In short, we perform a measurement of d$\mu$/d$n$ as a function of $n$ and $B$ (Extended Data Figure 2\textbf{a}). We then integrate d$\mu$/d$n$ across the entire density range for each magnetic field $B$ to obtain $\mu(n, B)$. $N$ points are evenly sampled across the density range. This discrete sampling procedure produces an energy spectrum $E(i, B)$, where $i \in [1, 2, ... , N]$. Plotting $E(i, B)$ for each $B$ produces the Hofstadter spectrum in Extended Data Figure 2\textbf{c}.

\subsection*{Estimation of $\theta_{23}$}
Nominally, the twist angle $\theta_{23}$ between layers 2 and 3 also creates a moiré superlattice. However, due the relatively large $\theta_{23}$, any features related to the filling of this secondary moiré superlattice are inaccessible within the experimentally available gate voltage range, making it challenging to unambiguously and precisely determine $\theta_{23}$ experimentally. However, we can estimate $\theta_{23}$ by measuring the renormalized Fermi velocity $v^*_F$ of the Dirac cone associated with layer 3.

Ignoring tunneling between layers 1 and 3, and assuming both the interlayer AA/BB and AB/BA hopping elements are equal to $t_0$, the renormalized Fermi velocity of layer 3, is given by
\begin{equation}
v^*_F = \frac{1-3\alpha^2_{23}}{1+6\alpha^2_{23}}v^0_F,
\end{equation}
where $\alpha_{23} = t_0/(\hbar v^0_F k_{\theta_{23}})$ and $k_{\theta} = 8 \pi \sin(\theta/2)/(3a)$~\cite{zhu2020twisted,uri2023superconductivity}. This gives a direct relation between $\theta_{23}$ and $v^*_F$:
\begin{equation}
\sin(\theta_{23}/2) = \frac{3 t_0 a}{8 \pi \hbar v^0_F} \sqrt{\frac{3(1+2v^*_F/v^0_F)}{1-v^*_F/v^0_F}}.
\end{equation}
We set $t_0 = 0.105$ eV and $v^0_F = 1\times10^6$ m/s, consistent with previous experimentally estimated values~\cite{uri2023superconductivity}. The above equation allows us to estimate $\theta_{23}$ by experimentally measuring $v^*_F$. The relationship between $\theta_{23}$ and $v^*_F$ is plotted in Extended Data Figure 2\textbf{d}.

At higher energies, the experimentally measured Hofstadter spectrum displays a set of dispersing features with enhanced density of states (DOS), which we identify as the Landau levels (LLs) of the Dirac cone in a finite magnetic field $B$ (Extended Data Figure 2\textbf{b}). We fit these states to the standard $\sqrt{B}$ field dependence of Dirac Landau levels:
\begin{equation}
E = E_0 + \mathrm{sgn}(N)v^*_F \sqrt{2 e \hbar N B},
\end{equation}
where $N$, $e$ and $\hbar$ are the LL index, electron charge and reduced Plank constant, respectively, and $E_0$ and $v^*_F$ are left as fitting parameters. We specifically fit to the most prominent observed LLs ($N= \pm1$, and $\pm2$) to estimate $v^*_F$. From this procedure and the data in Extended Data Figure 2 we find $v^*_F = 0.81 \times 10^6$ m/s, corresponding to $\theta_{23} \approx 3.45^\circ$. Using this method, we also estimate $\theta_{23} \approx 3.95^\circ$ and $\theta_{23} \approx 3.81^\circ$ at two other locations in the sample where $\theta_{12} = 1.14^\circ$ and $\theta_{12} = 1.20^\circ$, respectively. 
These values are consistent with the estimation of $\theta_{23} \sim 3$-$4^\circ$ from the relative twist angles between the flat edges of layers 2 and 3 seen in optical images during sample fabrication Extended Data Figure 6.
These locations are marked by the colored dots in Extended Data Figure 3\textbf{a}.
Lastly, we note that although the set up of our SET experiment necessarily introduces a small displacement field $D$ that cannot be independently controlled, the effect of this displacement field is negligible and does not meaningfully impact the extracted $v^*_F$~\cite{pierce2025tunable}.

\subsection*{Extraction of thermodynamic gaps}
The thermodynamic gap of an incompressible state is given by the corresponding step in the chemical potential $\Delta$, which is obtained by integrating the inverse compressibility d$\mu$/d$n$:
\begin{equation}
\Delta =  \mu(n_{+}) - \mu(n_{-}) = \int_{n_{-}}^{n_{+}} \left( \frac{{\rm{d}}\mu}{{\rm{d}}n} - \kappa^{-1}_B \right) \,\rm{d} \it{n}, 
\end{equation}
where $n_{+(-)}$ is the upper (lower) bound in density of the gapped state and $\kappa^{-1}_B$ is a small constant background that may need to be subtracted. We followed the same procedure as is described in detail in Ref.~\cite{yu2022correlated}.

\subsection*{Extraction of sawtooth compressibility strength}
The strength of the sawtooth in compressibility for the conduction (valence) band $\sigma_{e(h)}$ is calculated by the standard deviation of d$\mu$/d$n$ between $0.5 < \nu < 3.5$ ($-3.5 < \nu < -0.5$). Before calculating the standard deviation, a Savitzky–Golay filter is applied to the data to: 1) remove contributions from noise in the measurement and 2) remove contributions of the peak at $\nu=\pm2$ to the extracted standard deviation, thereby isolating only contributions related to the sawtooth compressibility itself. We find that the application of the Savitzky–Golay filter has no impact on the position or width of the peak in $\sigma_{e(h)}$ and simply shifts $\sigma_{e(h)}$ down by an overall constant (Extended Data Figure 4). Limiting the calculation of the sawtooth strength to $1.5<|\nu|<2.5$ to more closely match the region in which superconductivity is present does not meaningfully change its qualitative angle dependence.

\subsection*{Further discussion of $1\times1$ $\mu$m$^2$ spatial map}
Extended Data Figure 5 shows the systemic spatial dependence of $\Delta\mu_f = \Delta\mu_e + \Delta\mu_h$, $\Delta_2$, and $\sigma_{e(h)}$ in a $1\times1$ $\mu$m$^2$ area of the sample ($\Delta_{-2} = 0$ through out the entire area).
$\theta_{12}$ is the primary driver of the observed correlation physics. However, even at constant $\theta_{12}$, systematic spatial variability in each signature is evident, as seen Extended Data Figure 5\textbf{h}-\textbf{k}, where the color of each data point represent its $x$-coordinate in the grid. From these spatially-resolved measurements, we find that for a given $\theta_{12}$ both $\Delta\mu_f$ and $\Delta_2$ are suppressed from left (blue) to right (red), while $\sigma_{e}$ is instead enhanced. This highlights that the detailed microscopic conditions (beyond just $\theta_{12}$) that favor $\Delta_2$ or $\sigma_{e}$ differ.

The specific microscopic origins of the 2D spatial dependence in Extended Data Figure 5 is difficult to determine unambiguously from our experiment. More extensive theoretical modeling is needed to understand it fully, but that is beyond the scope of this work. We discuss two possible origins of the spatial variability observed in Extended Data Figure 5 even for constant $\theta_{12}$: local variations in strain, or changes in $\theta_{23}$. We speculate that the latter are the more likely explanation of the data than variations in strain, as we discuss further below. 

In MATBG, increased strain is predicted to enhance $\Delta\mu_f$, but suppress $\Delta_2$~\cite{yu_spin_2022}. Due to the relatively weak coupling between layers 2 and 3, a similar trend may be expected for this TTG device (though we are not aware of theoretical modeling for generic twist angles in TTG). In contrast to this expectation from strain, we instead observe the opposite trend: for constant $\theta_{12}$, we find in Extended Data Figure 5\textbf{h}-\textbf{i} that both $\Delta\mu_f$ and $\Delta_2$ generally decrease and increase together in space. Additionally, Extended Data Figure 1 shows that for constant $\theta_{12}$, increasing $\theta_{23}$ within the range of $\theta_{23}$ expected in our experiment tends to flatten the (single particle) bands further. Thus, although small spatial variability in strain is likely present, it is reasonable that variations in $\theta_{23}$ may be the dominant contributor to the observed spatial trends in Extended Data Figure 5. 

\subsection*{Extraction of the superconducting critical temperature}
In transport experiments in two-dimensional systems, three criteria are required to conclusively determine the presence of superconductivity: 1) zero longitudinal resistance $R_{xx}$, 2) non-linear d.c.~$I$-$V$ characteristics with zero voltage drop up to a critical current $I_c$, and 3) oscillations in the critical current as a function of perpendicular magnetic field that indicate phase-coherent transport. Our sample meets all three criteria (Fig.~\ref{fig:fig3}\textbf{d}-\textbf{f}), but verifying all three criteria at every $(n, D)$ point where superconductivity appears is impractical. 

To characterize superconductivity at every $(n, D)$ point, we measure $R_{xx}$ maps as a function of $n$ and $D$, and then vary the temperature $T$ in small, discrete steps. From this three-dimensional data set, we can extract the critical temperature $T_c$ as a function of $n$ and $D$ (Fig.~\ref{fig:fig3}\textbf{g}-\textbf{j}). 
We ascribe a finite $T_c$ only to  data points where $R_{xx}$ displays the characteristic superconducting $R_{xx} - T$ transition curve (Fig.~\ref{fig:fig3}\textbf{d}, for example) and where $R_{xx}$ falls below $100\ \Omega$. 
If any $(n, D)$ point does not meet both of these criteria we assign $T_c = 0$---either the sample is not superconducting at the specific $(n, D)$, or the superconductivity does not fully develop at base temperature. To determine $T_c$, we first linearly fit the $R_{xx} - T$ curves at high $T$, above the onset of the superconducting transition, and extrapolate to $T=0$. We define this to be the normal-state resistance $R_{xx}^\text{n}(T)$ as a function of $T$. We then take $T_c$ to be the temperature where 10\% of the extrapolated normal-state $R_{xx}^\text{n}(T)$ line intersects the measured $R_{xx} - T$ curve. Additionally, to estimate the uncertainty in the extracted $T_c$ from each $R_{xx}-T$ curve, we employed a ``leave-three-out'' jackknife approach. For each curve, all subsets with three points removed were fit individually, and $T_c$ was determined from each fit. The uncertainty is reported as the standard deviation of these $T_c$ values.

Lastly, while we plot the optimal (maximal) $T_c$ in Fig.~\ref{fig:fig5} at $V_t = 0$ from the line traces in Fig.~\ref{fig:fig4}, we note that the  qualitative shape and $\theta_{12}$ dependence of $T_c$ remains consistent independent of the details used to extract $T_c$ or $T^\mathrm{opt}_c$. This includes if we, for example, define $T_c$ to be where $R_{xx}$ reaches 20\% of the normal state resistance (as opposed to 10\%), if we extract $T^\mathrm{opt}_c$ at differed fixed values of $D$ (Extended Data Figure 8), or if we use the entire $n$-$D$ map to extract $T^\mathrm{opt}_c$ (rather than only considering a fixed $V_t$ or $D$). We note that the data points for $T^\mathrm{opt}_c$ plotted in Extended Data Figure 9 and Extended Data Figure 10 are the maximum $T_c$ extracted from the entire $n$-$D$ map.

\section*{Data availability}
Data that support the findings in this study are available at https://doi.org/10.5281/zenodo.18796678.

\setcounter{figure}{0}

\newpage

\begin{figure*}[p!]
    \renewcommand{\thefigure}{ED\arabic{figure}}
    \centering
    \includegraphics[width=1.99\columnwidth]{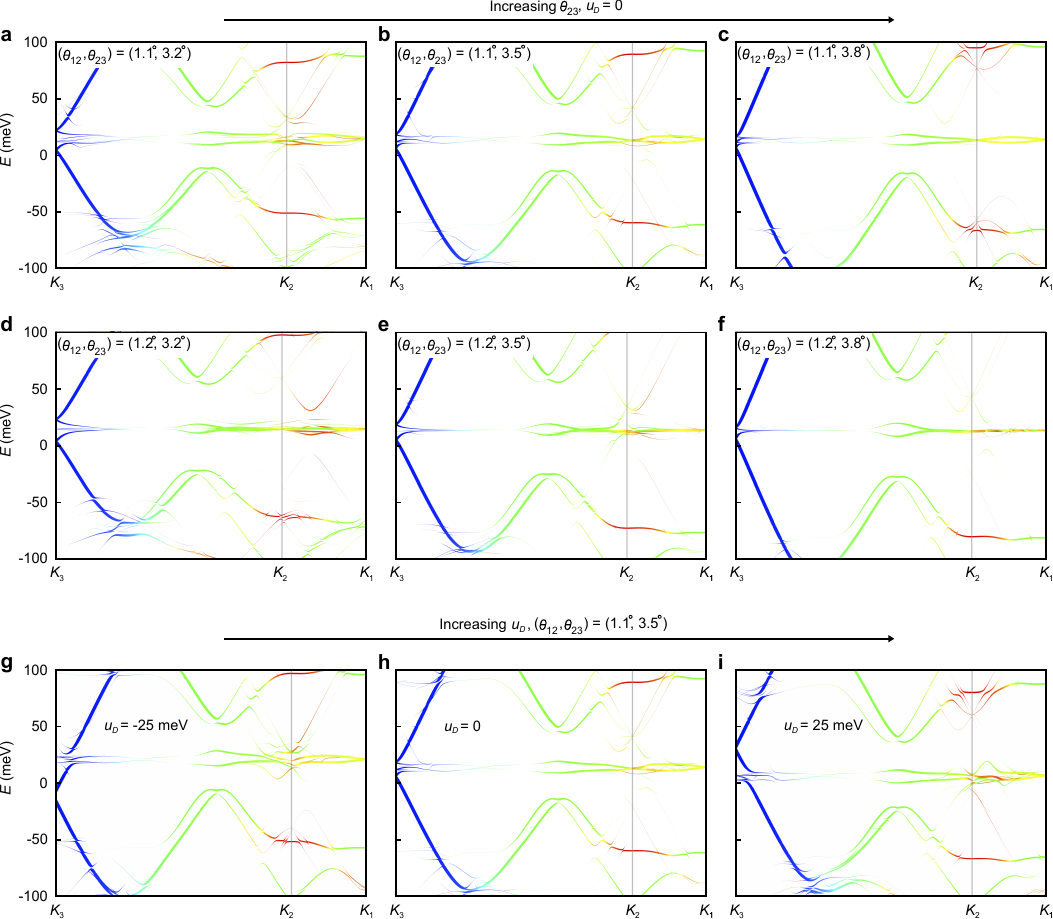}
    \caption{\textbf{Spectral functions of other ($\theta_{12}, \theta_{23}$) combinations and varying interlayer potentials}. \textbf{a}-\textbf{c}, Spectral function calculations for $\theta_{12}=1.1^\circ$ and different $\theta_{23}$ along a path between the $K$ points for each layer. \textbf{d}-\textbf{f}, Same as \textbf{a}-\textbf{c}, but for $\theta_{12}=1.2^\circ$. Colors reflect layer character, as defined in the inset of Fig.~\ref{fig:fig1}\textbf{b}. Each calculation is taken at zero displacement field. \textbf{g}-\textbf{i}, Spectral function calculations for ($\theta_{12}, \theta_{23}$) = ($1.1^\circ, 3.5^\circ$) for different interlayer potential differences. The potentials for layers 1, 2, and 3 are set to $-u_D$, $0$, and $u_D$, respectively, to simulate the effect of a displacement field $D$ (Methods). These calculations demonstrate that the primary effect of the displacement field is to tune the relative energies of the flat and Dirac-like bands. Colors reflect layer character, as defined in the inset of Fig.~\ref{fig:fig1}\textbf{b}.
    }
    \label{fig:sf}
\end{figure*}

\begin{figure*}[p!]
    \renewcommand{\thefigure}{ED\arabic{figure}}
    \centering
    \includegraphics[width=1.7\columnwidth]{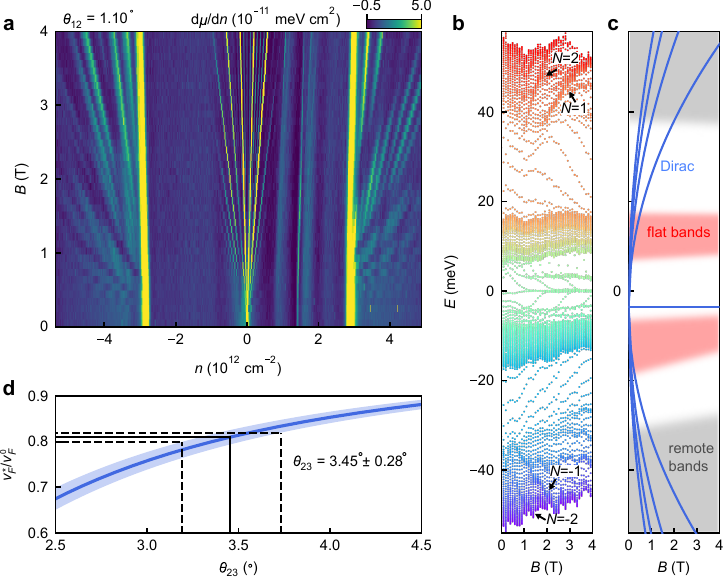}
    \caption{\textbf{Estimation of $\theta_{23}$}. \textbf{a}, Landau fan measurement of d$\mu$/d$n$ as a function of $n$ and $B$. \textbf{b}, Measurement of the many-body Hofstadter spectrum for $\theta_{12} = 1.10^\circ$, extracted from the data in \textbf{a} (Methods). The density of dots reflects the many-body density of states (DOS) and the different colors correspond to states at different $n$. The enhanced DOS from the $N=\pm1, \pm2$ LLs from the Dirac-like cone are indicated with arrows. We fit a Fermi velocity of $v^*_F = (8.1 \pm 0.1) \times 10^5$ m/s to the LLs from the Dirac-like cone. \textbf{c}, Schematic illustration of the origin of different features in panel \textbf{b}. \textbf{d}, Renormalized Fermi velocity $v^*_F$ as a function of $\theta_{23}$. The horizontal black line corresponds to $v^*_F = 8.1 \times 10^5$ m/s, while the horizontal dashed black lines correspond to $v^*_F = (8.1 \pm 0.1) \times 10^5$ m/s. Assuming an uncertainty in $t_0$ of $t_0 =  0.105 \pm .005$ eV (shaded light blue area), this corresponds to a twist angle $\theta_{23} = 3.45^\circ$ (vertical black line) with an uncertainty of $\pm 0.28^\circ$ (vertical black dashed lines). 
} 
    \label{fig:theta23}
\end{figure*}

\begin{figure*}[p!]
    \renewcommand{\thefigure}{ED\arabic{figure}}
    \centering
    \includegraphics[width=1.7\columnwidth]{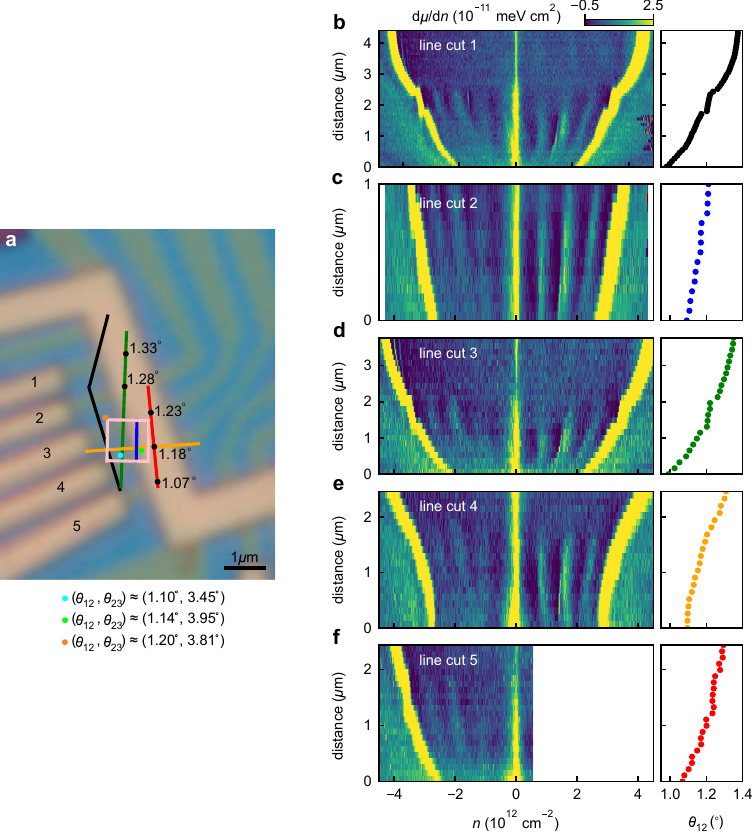}
    \caption{\textbf{Additional spatial line cuts of d$\mu$/d$n$}. \textbf{a}, Trajectories of spatial line cuts of d$\mu$/d$n$ overlaid on the Hall bar. The black trajectory is the line cut plotted in Fig.~\ref{fig:fig2}. The pink square corresponds to the location of the $1 \times 1$ $\mu$m$^2$ grid shown in Fig.~\ref{fig:grid}. The colored dots inside the pink square denote locations where $\theta_{23}$ is estimated via the method described in Fig.~\ref{fig:theta23}. At the locations where $\theta_{12} = 1.10^\circ$, $1.14^\circ$, and $1.20^\circ$, we estimated $\theta_{23} \approx 3.45^\circ \pm 0.28^\circ$, $3.95^\circ \pm 0.33^\circ$, and $3.81^\circ \pm 0.32^\circ$, respectively. \textbf{b}, d$\mu$/d$n$ along the black line in \textbf{a}, from bottom to top. Data are identical to those in Fig.~\ref{fig:fig2}\textbf{a}, but are reproduced for clarity and comparison. \textbf{c}, Line cut of d$\mu$/d$n$ along the blue line in \textbf{a}, from bottom to top. \textbf{d}, Line cut of d$\mu$/d$n$ along the green line in \textbf{a}, from bottom to top. \textbf{e}, Line cut of d$\mu$/d$n$ along the orange line in \textbf{a}, from left to right. \textbf{f}, Line cut of d$\mu$/d$n$ along the red line in \textbf{a}, from bottom to top. Each spatial line cut is measured at $T=1.6$~K.
} 
    \label{fig:linecuts}
\end{figure*}

\begin{figure*}[p!]
    \renewcommand{\thefigure}{ED\arabic{figure}}
    \centering
    \includegraphics[width=1.4\columnwidth]{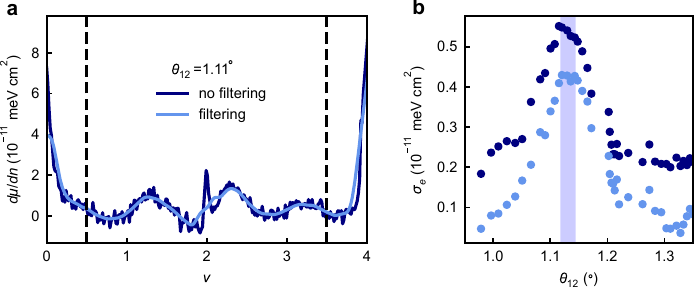}
    \caption{\textbf{Effect of filtering on the the strength of the sawtooth in compressibility}. \textbf{a}, Example trace of d$\mu$/d$n$ as a function of $\nu$ for $\theta_{12}=1.11^\circ$ with (light blue) and without (navy) Savitzky–Golay filtering. Black dashed lines indicate the bounds between which $\sigma_e$ is calculated. \textbf{b}, $\sigma_e$ as a function of $\theta_{12}$ from the line cut in Fig.~\ref{fig:fig2}\textbf{a} with and without filtering. The filtered and unfiltered data sets agree qualitatively, differing only by an approximately constant vertical offset.  
} 
    \label{fig:filter}
\end{figure*}

\begin{figure*}[p!]
    \renewcommand{\thefigure}{ED\arabic{figure}}
    \centering
    \includegraphics[width=1.99\columnwidth]{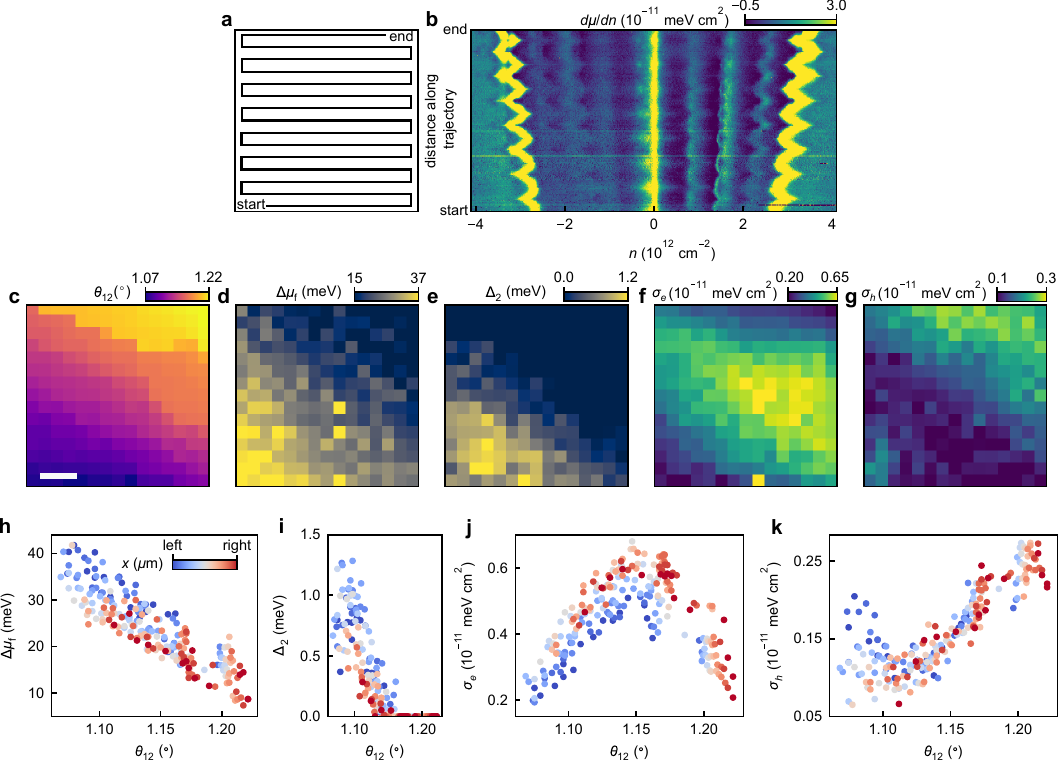}
    \caption{\textbf{d$\mu$/d$n$ within a $1 \times 1$ $\mu$m$^2$ area}. \textbf{a}, Rasterized trajectory of the SET tip during the measurement. \textbf{b}, d$\mu$/d$n$ as a function of $n$ along the trajectory shown in \textbf{a}. \textbf{c}-\textbf{g}, Spatial map of $\theta_{12}$, $\Delta \mu_f$ = $\Delta \mu_e$ + $\Delta \mu_h$, $\Delta_2$, $\sigma_e$, and $\sigma_h$, respectively, within the $1 \times 1$ $\mu$m$^2$ grid area. Each is extracted from the data in \textbf{b}. Note that there is no correlated state at $\nu=-2$ observed within the grid, i.e.~$\Delta_{-2} = 0$ throughout this area. \textbf{h}-\textbf{k}, Scatter plots of $\Delta \mu_f$, $\Delta_2$, $\sigma_e$, and $\sigma_h$, respectively, as a function of $\theta_{12}$. The color of each data point indicates its $x$-coordinate within the $1 \times 1$ $\mu$m$^2$ area. 
} 
    \label{fig:grid}
\end{figure*}

\begin{figure*}[p!]
    \renewcommand{\thefigure}{ED\arabic{figure}}
    \centering
    \includegraphics[width=1.4\columnwidth]{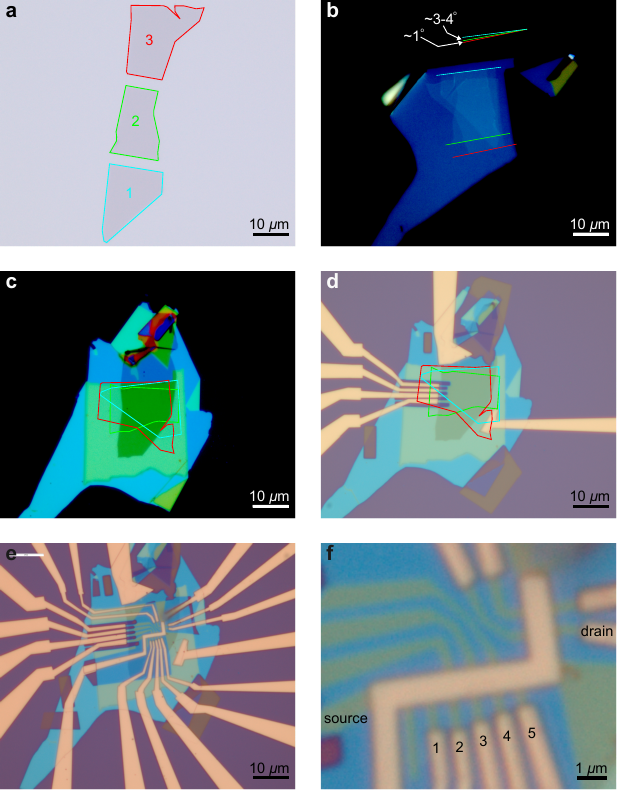}
    \caption{\textbf{Device fabrication}. \textbf{a}, Optical image of the graphene flakes used in the TTG device after being cut into three pieces with an atomic force microscope (AFM). \textbf{b}, The trilayer stack after being picked up with an hBN flake on the polymer stamp, but before deposition onto the bottom gate. Colored lines indicate AFM cut edges, from which we estimate $\theta_{12} \sim 1^{\circ}$ and $\theta_{23} \sim 3^{\circ}-4^{\circ}$. \textbf{c}, The final stack immediately after deposition onto a pre-stacked hBN and few-layer thick graphite gate on a Si/SiO$_2$ substrate. \textbf{d}, Image of the device after etching and metallization, at which point it was measured with the scanning SET. \textbf{e}, Image of the device after the scanning SET experiments were completed and after additional etching and metallization into a Hall bar geometry with an added Cr/Au top gate. \textbf{f}, Zoomed in image of the Hall bar device. Contacts 1-5 were used as voltage leads for four-probe resistance measurements. The source and drain are also labeled.
} 
    \label{fig:fab}
\end{figure*}

\begin{figure*}[p!]
    \renewcommand{\thefigure}{ED\arabic{figure}}
    \centering
    \includegraphics[width=1.8\columnwidth]{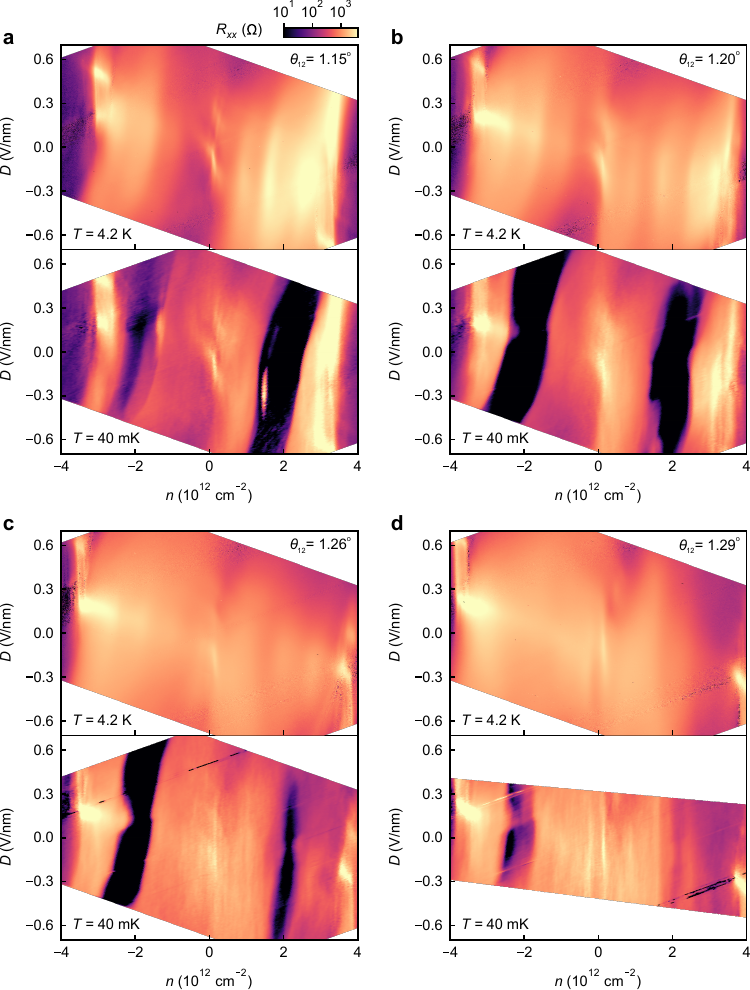}
    \caption{\textbf{Transport measurements at different temperatures for each contact pair}. \textbf{a}, Upper panel: longitudinal resistance $R_{xx}$ as function of $n$ and $D$ for contacts 1 and 2 at $T = 4.2$ K. Lower panel: $R_{xx}$ as a function of $n$ and $D$ for contacts 1 and 2 at $T = 40$ mK. \textbf{b}-\textbf{d}, Same as \textbf{a}, but for the other contact pairs.
} 
    \label{fig:transport}
\end{figure*}

\begin{figure*}[p!]
    \renewcommand{\thefigure}{ED\arabic{figure}}
    \centering
    \includegraphics[width=1.7\columnwidth]{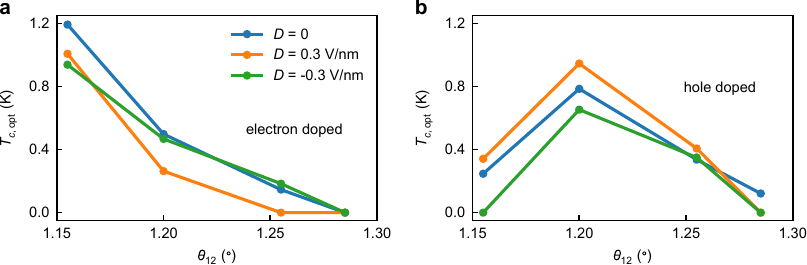}
    \caption{\textbf{Critical temperature at different displacement fields}. \textbf{a}, Critical temperature at optimal doping $T^{\mathrm{opt}}_c$ for the electron-doped superconductor as a function of $\theta_{12}$ for three different displacement fields. \textbf{b}, Same as \textbf{a}, but for hole-doping.
} 
    \label{fig:Tc_diffD}
\end{figure*}

\begin{figure*}[p!]
    \renewcommand{\thefigure}{ED\arabic{figure}}
    \centering
    \includegraphics[width=1.8\columnwidth]{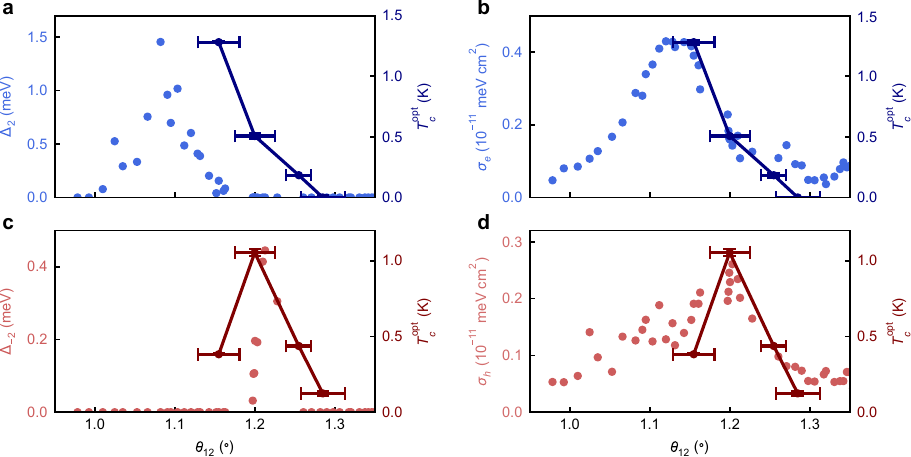}
    \caption{\textbf{Comparison of compressibility data from the first line cut with transport}. \textbf{a}, Gap $\Delta_2$ of the $\nu=2$ correlated insulator (left axis) as measured from the line cut in Fig.~\ref{fig:fig2} (same as the line cut 1 in Fig.~\ref{fig:linecuts}) and critical temperature at optimal doping and displacement field $T^{\mathrm{opt}}_c$ (right axis) as measured from transport as a function of twist angle $\theta_{12}$. \textbf{b}, Strength of the sawtooth compressibility $\sigma_e$ for the conduction band (left axis) as measured from the line cut and $T^{\mathrm{opt}}_c$ (right axis) as measured from transport as a function of twist angle $\theta_{12}$. \textbf{c}-\textbf{d}, Same as \textbf{a}-\textbf{b}, but for hole doping. 
} 
    \label{fig:Tc_linecut}
\end{figure*}

\begin{figure*}[p!]
    \renewcommand{\thefigure}{ED\arabic{figure}}
    \centering
    \includegraphics[width=1.8\columnwidth]{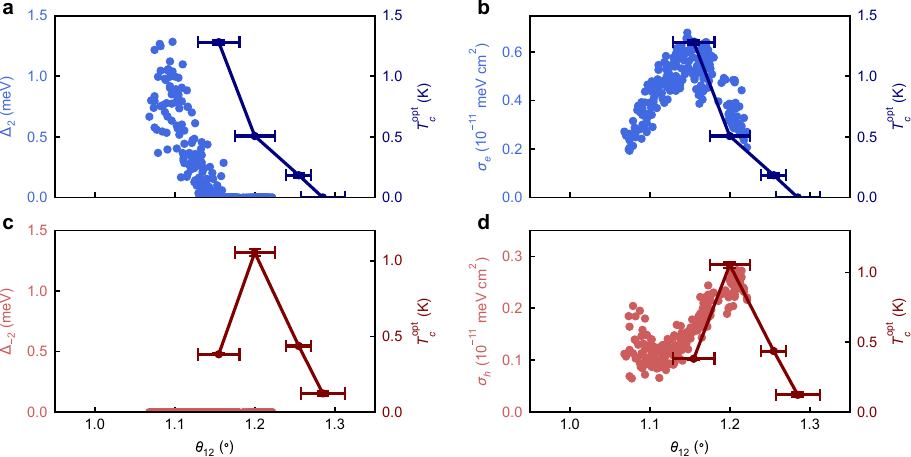}
    \caption{\textbf{Comparison of compressibility data from the $1 \times 1$ $\mu$m$^2$ grid with transport}. \textbf{a}, $\Delta_2$, measured within the grid (Fig.~\ref{fig:grid}; left axis) and critical temperature at optimal doping and displacement field $T^{\mathrm{opt}}_c$ determined from transport measurements (right axis) as a function of $\theta_{12}$. \textbf{b}, $\sigma_e$ measured within the grid (left axis) and optimal $T_c^{\textrm{opt}}$ determined from transport measurements (right axis) as a function of $\theta_{12}$.
    \textbf{c}-\textbf{d}, Same as \textbf{a}-\textbf{b}, but for the hole doping.
} 
    \label{fig:Tc_grid}
\end{figure*}

\end{document}